\pdfoutput=1
\documentclass[%
 reprint,
superscriptaddress,
preprintnumbers,
bibnotes,
tightenlines,
notitlepage,
longbibliography,
twocolumn,
 amsmath,amssymb,
 aps,
]{revtex4-1}

\usepackage{graphicx}

\usepackage{dcolumn}
\usepackage{bm}
\usepackage[usenames,dvipsnames,svgnames,table]{xcolor}
\usepackage[colorlinks = true,
            linkcolor = Navy,
            urlcolor  = Navy,
            citecolor = Navy,
            anchorcolor = Navy]{hyperref}
\usepackage[caption=false]{subfig}

\usepackage{feynmp-auto}

\usepackage{siunitx}

\usepackage{lineno}
\usepackage{placeins}

\usepackage{makecell}

\usepackage{booktabs}
\newcolumntype{C}{>{$}c<{$}}
\AtBeginDocument{
\heavyrulewidth=.08em
\lightrulewidth=.05em
\cmidrulewidth=.03em
\belowrulesep=.65ex
\belowbottomsep=0pt
\aboverulesep=.4ex
\abovetopsep=0pt
\cmidrulesep=\doublerulesep
\cmidrulekern=.5em
\defaultaddspace=.5em
}

\usepackage[utf8]{inputenc}

\begin{document}

\title{Strategy to measure tau $g-2$ via photon fusion in LHC proton collisions }% 

\author{Lydia Beresford}
 \email{lydia.beresford@desy.de}
 \affiliation{Deutsches Elektronen-Synchrotron DESY, Notkestr.\ 85, 22607 Hamburg, Germany}
 \author{Savannah Clawson}
 \email{savannah.clawson@desy.de}
 \affiliation{Deutsches Elektronen-Synchrotron DESY, Notkestr.\ 85, 22607 Hamburg, Germany} 
 \author{Jesse Liu}
 \email{jesseliu@hep.phy.cam.ac.uk}
 \affiliation{Cavendish Laboratory, University of Cambridge, Cambridge CB3 0HE, UK} 

% Institutional report number
\preprint{DESY-24-039}

%\date{\today}

\begin{abstract}
Measuring the tau-lepton ($\tau$) anomalous magnetic moment $a_\tau = (g_\tau -2)/2$ 
in photon fusion production ($\gamma\gamma\to\tau\tau$) tests foundational Standard Model principles. 
However, $\gamma\gamma\to\tau\tau$ eludes observation in LHC proton collisions (pp) despite enhanced new physics sensitivity from higher-mass reach than existing probes.
We propose a novel strategy to measure $\text{pp} \to \text{p}(\gamma\gamma\to\tau\tau)\text{p}$ by introducing the overlooked electron-muon signature with vertex isolation for signal extraction. 
Applying the effective field theory of dipole moments, we estimate 95\% CL sensitivity of $-0.0092 < a_\tau < 0.011$ assuming 300~fb$^{-1}$ luminosity and 5\% systematics. 
This fourfold improvement beyond existing constraints opens a crucial path to unveiling new physics imprinted in tau-lepton dipoles.
\end{abstract}

\maketitle

%-------------------------------------
\section{\label{sec:intro} Introduction}
%-------------------------------------

Precise measurements of electromagnetic (EM) dipoles are fundamental tests of the Standard Model (SM) that could reveal beyond-the-SM (BSM) physics. 
A cornerstone SM principle is lepton universality, where all three generations (electron $e$, muon $\mu$, tau-lepton $\tau$) couple equally to gauge bosons. 
The leading SM loop correction from quantum fluctuations is also flavor universal, shifting magnetic moments by the Schwinger term $\alpha_\text{EM} / 2\pi \simeq 0.0012$~\cite{Schwinger:1948iu,Kusch:1948mvb}. 
The electron and muon anomalous magnetic moments $a_{e,\mu} = (g_{e,\mu}-2)/2$ are now tested to 13~\cite{PhysRevLett.97.030801,Hanneke:2010au,PhysRevLett.106.080801,Aoyama:2012wj,Parker191,Aoyama:2012wk,Keshavarzi:2018mgv,Davier:2019can,Morel:2020dww,Fan:2022eto} and 10 decimal places~\cite{Bennett:2006fi,Muong-2:2021ojo,Muong-2:2023cdq,Aoyama:2020ynm}, respectively. 
However, the tau-lepton counterpart $a_{\tau}$ is still compatible with zero to two decimal places~\cite{Tanabashi:2018oca} as its 0.3~ps proper lifetime~\cite{ALEPH:1997roz,DELPHI:2003zcz,L3:2000dya,Belle:2013teo} precludes storage-ring probes~\cite{Muong-2:2023cdq}. 
The existence of tau-lepton loop interactions with photons in nature thus remains untested.

The most precise single-experiment $a_{\tau}$ constraint is a $-0.052 < a_\tau^\text{obs} < 0.013 $ 95\% CL limit by DELPHI~\cite{Abdallah:2003xd} at the Large Electron Positron Collider (LEP), with similar precision by L3 and OPAL~\cite{1998169,Ackerstaff:1998mt}. 
ATLAS and CMS recently pioneered Large Hadron Collider (LHC) probes of $a_{\tau}$ using photon fusion production of tau-leptons ($\gamma\gamma\to\tau\tau$) in lead-lead (PbPb) data~\cite{ATLAS:2022ryk,CMS:2022arf}; the ATLAS 95\% CL limit is $-0.057 < a_\tau^\text{obs} < 0.024$. 
Such large experimental uncertainties relative to the SM prediction $a_{\tau,\,\text{SM}}^\text{pred} = 0.001\,177\,21\,(5)$~\cite{Eidelman:2007sb} could conceal BSM dynamics, for example those motivated by lepton sector tensions~\cite{Fukuyama:2005bh,Dutta:2018fge,Davoudiasl:2018fbb,Bauer:2019gfk,Endo:2019bcj,LHCb:2015gmp,LHCb:2023zxo,Abdesselam:2019dgh,Allanach:2015gkd,DiChiara:2017cjq,Biswas:2019twf,Yin:2016shg,Yamaguchi:2016oqz,Babu:2018qca,Iguro:2023rom,Allanach:2022iod,Allanach:2023uxz}. 
Specific models predict quadratic scaling $\delta a_{\ell} \propto m_{\ell}^2$ with lepton mass $m_\ell$~\cite{Moroi:1995yh,Ibrahim:1999hh,Martin:2001st}, implying $(m_\tau /m_\mu)^2 \simeq 280$ times larger effects for $a_\tau$ than $a_\mu$.
New physics can also violate charge-parity (CP) symmetry, inducing an electric dipole $d_{\tau}$. 
Standard LHC proton-proton (pp) collisions reach higher $\mathcal{O}\text{(TeV)}$ masses, enhancing BSM dipole sensitivity over $\mathcal{O}\text{(100~GeV)}$ in PbPb~\cite{DELAGUILA1991256,Beresford:2019gww,Dyndal:2020yen,Burmasov:2023cwv,Goncalves:2020btj,Verducci:2023cgx}. 
Despite this key benefit, cross-section yielding over 30 million events to date, and major photon-fusion advances~\cite{Piotrzkowski:2000rx,Baltz:2007kq,Albrow:2008pn,deFavereaudeJeneret:2009db,dEnterria:2009cwl,ATLAS:2015wnx,Aaboud:2017oiq,Aaboud:2016dkv,Chatrchyan:2012tv,CMS:2013hdf,Khachatryan:2016mud,ATLAS:2020hii,ATLAS:2020epq,ATLAS:2022srr,dEnterria:2013zqi,Aaboud:2017bwk,Sirunyan:2018fhl,Aad:2019ock,Howarth:2020uaa,Goncalves:2020saa,Chapon:2009hh,Fichet:2013gsa,Ellis:2017edi,Knapen:2016moh,Baldenegro:2018hng,Ohnemus:1993qw,Schul:2008sr,HarlandLang:2011ih,Beresford:2018pbt,Harland-Lang:2018hmi,Bruce:2018yzs,dEnterria:2023npy,ATLAS:2023zfc,TOTEM:2021zxa,CMS:2022dmc}, 
$\gamma\gamma \to \tau\tau$ remarkably evades observation in pp data. 

\begin{figure}
    \centering
    \includegraphics[width=0.25\textwidth]{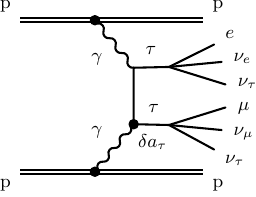}%
    \includegraphics[width=0.22\textwidth]{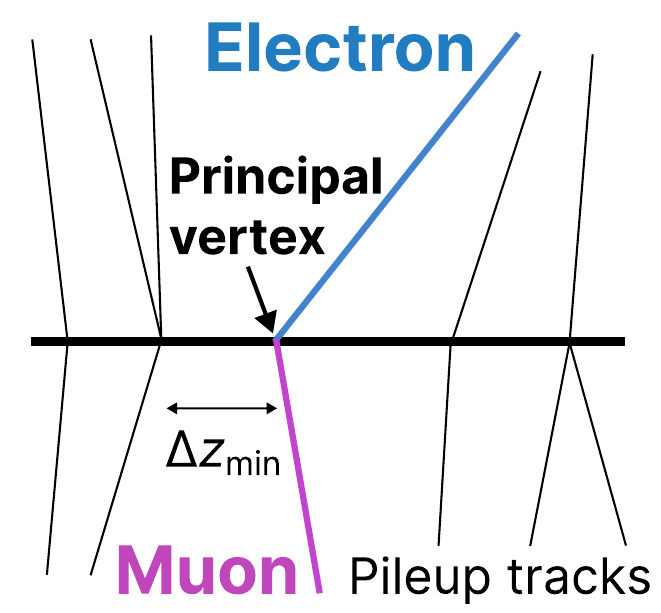}
    \caption{\label{fig:feynGraphs_yy2tautau} 
    Tau-leptons produced from photon fusion in proton beams with electron-muon $\tau\tau \to e\nu\nu \mu\nu\nu$ decays as a Feynman diagram (left) and detector signature illustrating the vertex isolation technique for the electron-muon vs pileup tracks (right). 
    New physics can modify the magnetic moment $\delta a_\tau$.
    }   
\end{figure}

This paper proposes the strategy to measure $\gamma\gamma\to\tau\tau$ in the fully-leptonic channel and tau-lepton EM dipoles in LHC pp collisions (Fig.~\ref{fig:feynGraphs_yy2tautau}).
We initiate the first Monte Carlo (MC) simulation analysis of the $\text{pp} \to \text{p}(\gamma\gamma \to \tau\tau)\text{p}$ signal that includes important weak-boson backgrounds and detector effects neglected in earlier work~\cite{Atag:2010ja}.
Hadronic channels have the highest signal rates, but higher-threshold triggers, estimating jet mis-tag rates, and multiple pp interactions (pileup) present greater challenges. 
We overcome these longstanding obstacles by leveraging recent progress~\cite{ATLAS:2020mve,ATLAS:2020iwi,ATL-PHYS-PUB-2021-026} to introduce the electron-muon signature, track-vertex isolation techniques (Fig.~\ref{fig:feynGraphs_yy2tautau}, right), and kinematic discriminants all unexplored for pp probes of $\gamma\gamma\to\tau\tau$. 
We exploit high-momentum kinematics unique to pp events that augment BSM dipole sensitivity.
We also propose critical strategies for controlling systematics.
Our proposal complements other production modes~\cite{Samuel1994,Hayreter:2013vna,Hayreter:2013vna,Galon:2016ngp,Fomin:2018ybj,Fu:2019utm,Haisch:2023upo} and future
facilities~\cite{Eidelman:2016aih,Crivellin:2021spu,Koksal:2018env,PhysRevD.89.037301,Rajaraman:2018uyb,Koksal:2018xyi,Gutierrez-Rodriguez:2019umw,Chwastowski:2022fzk,Fael:2013ij,Koksal:2021gyd,Chen:2018cxt,Tran:2020tsj,Koksal:2018vtt,Qian:2022owu}, while broadening the precision tau-lepton~\cite{ALEPH:2001uca,ATLAS:2017xuc,CMS:2023mgq,ATLAS:2020xea,BaBar:2009qmj,BESIII:2014srs,Belle-II:2023izd} and search programs~\cite{Alves:2011wf,Abdallah:2015ter,Barr:2016sho,ATLAS:2015nsi,ATLAS:2019itm,ATLAS:2019lng,ATLAS:2019gti,ATLAS:2020zms,CMS:2018eqb,CMS:2019zmn,CMS:2022qva,LHCb:2019vmc}.

%-------------------------------------
\section{\label{sec:sim_det} Model and simulation}
%-------------------------------------

Relativistic field theory generalizes the Schr\"{o}dinger-Pauli Hamiltonian $\mathcal{H} = -\boldsymbol{\mu}_\tau\cdot \mathbf{B} - \mathbf{d}_\tau \cdot \mathbf{E}$ describing EM dipoles into an effective Lagrangian coupling the Dirac spinor tensor $\sigma^{\mu\nu} = \mathrm{i}[\gamma^\mu, \gamma^\nu]/2$ to the photon field $F_{\mu\nu}$ 
\begin{linenomath*}
\begin{align}
    \mathcal{L}_\text{dipole} = \tfrac{1}{2} \bar{\tau}_\text{L}\sigma^{\mu\nu}  \left(a_\tau \tfrac{e}{2m_\tau} - \mathrm{i} d_\tau \gamma_5 \right) \tau_\text{R} F_{\mu\nu},
\end{align}
\end{linenomath*}
where $\tau_\text{L\,(R)}$ is the left- (right-) handed tau-lepton spinor and $\gamma^5$ satisfies the $\{\gamma^5, \gamma^\mu\} = 0$ anticommutator. 
Quantum electrodynamics (QED) predicts a vanishing electric dipole $d_\tau = 0$ and vacuum fluctuations induce $a_\tau \neq 0$.

\begin{figure}
    \centering
    \includegraphics[width=0.48\textwidth]{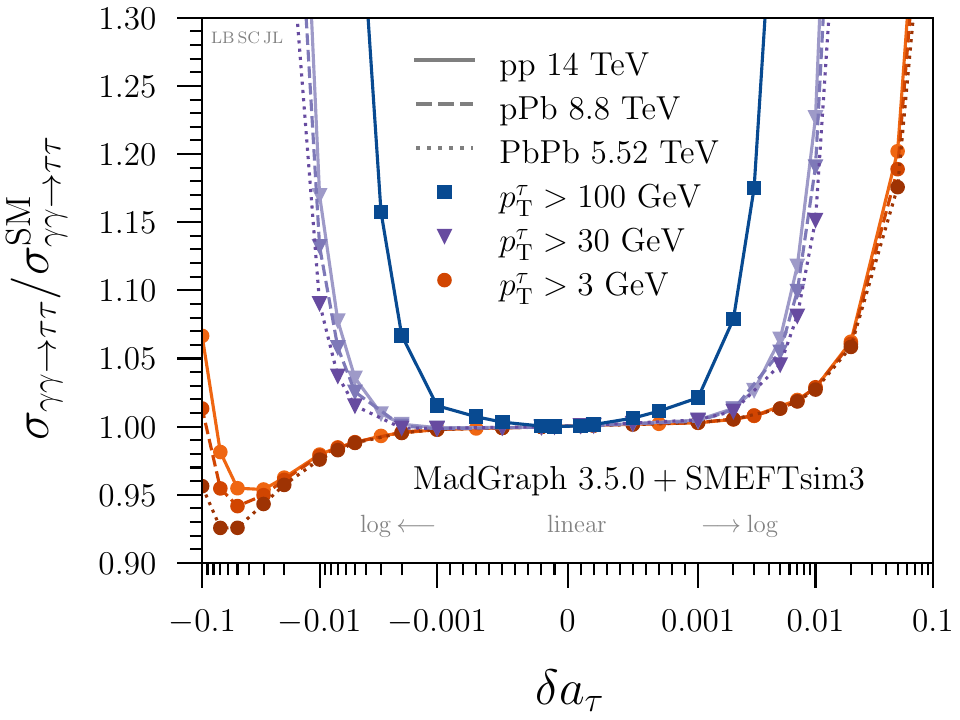}
    \caption{Cross-sections relative to the SM $\sigma_{\gamma\gamma\to\tau\tau}/\sigma_{\gamma\gamma\to\tau\tau}^\text{SM}$ for  elastic $\gamma\gamma\to\tau\tau$ production vs magnetic moment variations $\delta a_\tau$ with proton-proton pp 14~TeV (solid), proton-lead pPb 8.8~TeV (dashed), lead-lead PbPb 5.52~TeV (dotted) beams given tau-lepton transverse momentum $p_\text{T}^\tau >100$~GeV (squares, only pp), 30~GeV (triangles), 3~GeV (circles).}
    \label{fig:aa2tautau_xsec}
\end{figure}

New physics at high-mass scale $\Lambda$ can shift EM dipoles
beyond SM values at low momentum transfers $q$ satisfying $q^2 \ll \Lambda^2$, parameterized by SM Effective Field Theory (SMEFT)~\cite{Escribano:1993pq}.
Following Ref.~\cite{Beresford:2019gww}, we introduce the dimension-six SMEFT operator~\footnote{The $\tau$-lepton dipole moments can also be modified by a second dimension-six SMEFT operator, $C_{\tau W}$, which is set this to zero without loss of generality.}
\begin{linenomath*}
\begin{align}
    \mathcal{L}_\text{SMEFT} = \left(C_{\tau B} /\Lambda^2\right) \bar{L}_{\tau}\sigma^{\mu\nu} \tau_R H B_{\mu\nu}
    \label{eq:BSMLagrangian}
\end{align}
\end{linenomath*}
that modifies $(a_\tau, d_\tau)$ at tree level~\cite{Eidelman:2016aih}, where $L_\tau$ ($H$) is the tau-lepton (Higgs) doublet, $B_{\mu\nu}$ is the hypercharge field, and $C_{\tau B}$ is a dimensionless, complex Wilson coefficient in the Warsaw basis~\cite{Grzadkowski:2010es}. 
We implement Eq.~\eqref{eq:BSMLagrangian} in \textsc{FeynRules}~\cite{Alloul:2013bka} using the \texttt{SMEFTsim\_general\_alphaScheme\_UFO} model in \textsc{SMEFTsim}~3.0.2~\cite{Brivio:2017btx,Brivio:2020onw}.
The real (\texttt{ceBRe33}) and imaginary (\texttt{ceBIm33}) parts of $C_{\tau B}$ in the \texttt{restrict\_SMlimit\_massless} parameter card map to dipole shifts
\begin{linenomath*}
\begin{align}
    \delta a_\tau &= \frac{2m_\tau}{e}\frac{\mathrm{Re}\left[C_{\tau B}\right]}{M},\quad
    \delta d_\tau = \frac{\mathrm{Im}\left[C_{\tau B}\right]}{M},
    \label{eq:delta_a_d_tau_defn}
\end{align}
\end{linenomath*}
defining $M = \Lambda^2/ (\sqrt{2}v\cos\theta_W)$ in terms of the Weinberg angle $\theta_W$ and $v = 246$~GeV, $m_\tau = 1.776$~GeV, and $e = 1/\sqrt{4\pi}$ in Heaviside-Lorentz units. 
We interface \textsc{SMEFTsim} with \textsc{MadGraph}~3.5.0~\cite{Alwall:2011uj,Alwall:2014hca} for cross-section calculation and MC event simulation.

Canonical calculations of LHC photon-fusion cross-sections $\sigma_{\gamma\gamma \to \tau\tau}^{(\text{pp})}$ factorize into convolutions of the elementary cross-section $\hat{\sigma}_{\gamma\gamma \to \tau\tau}$ with the photon flux $n(x)$~\cite{Budnev:1974de}
\begin{linenomath*}  
\begin{equation}
    \sigma_{\gamma\gamma \to \tau\tau}^{(\text{pp})} = \int \mathrm{d} x_1 \mathrm{d} x_2 \,
    n(x_1) n(x_2)\,
    \hat{\sigma}_{\gamma\gamma \to \tau\tau},
    \label{eq:xsec_excl_photon}
\end{equation}
\end{linenomath*}
where $x_i = E_i / E_\text{beam}$ is the photon energy $E_i$ emitted from proton $i$ normalized to beam energy $E_\text{beam}$.  
We adopt the charge form factor (\texttt{ChFF}) flux from \textsc{gamma-UPC}~1.0~\cite{Shao:2022cly}, which includes nonfactorizable soft-survival corrections to Eq.~\eqref{eq:xsec_excl_photon}. 
The cross-section is proportional to the amplitude squared $\hat{\sigma}_{\gamma\gamma\to\tau\tau} \propto |\mathcal{A}|^2$, 
\begin{linenomath*}  
\begin{align}
    |\mathcal{A}|^2 = |\mathcal{A}_\text{SM}|^2 + 2\mathrm{Re}\left(\mathcal{A}_\text{SM}\mathcal{A}_\text{BSM}^*\right) + |\mathcal{A}_\text{BSM}|^2.
    \label{eq:matrix_element}
\end{align}
\end{linenomath*}
Generating $\gamma\gamma \to \tau\tau$ events with up to two EFT vertices models this SM-BSM quantum interference.
The correspondence to static EM dipoles formally applies in the real photon limit $q^2_\gamma \to 0$, satisfied for elastic photons from LHC protons $q^2_\gamma/m_\tau^2 \simeq (280~\text{MeV}/1776~\text{MeV})^2 \ll 1$.

We find the 14~TeV cross-section $\sigma_{\gamma\gamma\to\tau\tau}^\text{(pp)}$ for elastic $\gamma\gamma \to \tau\tau$ production is 150~pb, yielding $4.5\times 10^7$ events at 300~fb$^{-1}$ luminosity. 
Imposing transverse momentum $p_\text{T}^{\tau} > 30~(100)$~GeV, the 100 (2.5)~fb cross-section yields favorable $3\times 10^4$ (750) events at 300~fb$^{-1}$, whereas the 25~nb (5.1~pb) PbPb cross-section yields fewer (negligible) 100~(0.02) events at 4~nb$^{-1}$.
Figure~\ref{fig:aa2tautau_xsec} shows the relative cross-section variations versus $\delta a_\tau$, assuming $\delta d_\tau = 0$ for different beams and minimum $p_\text{T}^{\tau}$.
Requiring $p_\text{T}^\tau > 30~(100)$~GeV modifies $\sigma_{\gamma\gamma\to\tau\tau}^\text{(pp)}$ by a measurable 5\% for per-mille $\delta a_\tau = 0.005~(0.001)$ shifts, dramatically improving $\delta a_\tau$ sensitivity when probing scales only accessible to pp beams.
We generate around 2 million elastic $\gamma\gamma\to\tau\tau$ events per point for 27 coupling variations in the range $\delta a_\tau \in [-0.015, 0.015]$ with $\delta d_\tau = 0$, requiring $p_\text{T}^{\tau} > 15$ GeV in \textsc{MadGraph} and fully leptonic decays using \textsc{Pythia}~8.306~\cite{Sjostrand:2007gs,Bierlich:2022pfr} customized for photon fusion~\cite{ippp-workshop2023,ATLAS:2020iwi} to improve MC statistics.
Dissociative photon-fusion (indicated by p$^{*}$) and background simulation follows standard \textsc{MadGraph}+\textsc{Pythia} procedures~\cite{Alwall:2014hca,Ball:2012cx,Mangano:2006rw} detailed in Appendix~\ref{apndx:MCdetails}, generating at least 1 million events per process.
All samples have \textsc{Delphes} 3.5.0~\cite{deFavereau:2013fsa} detector emulation applied.
We assume 300~fb$^{-1}$ for the 13--13.6~TeV dataset and 4000~fb$^{-1}$ for the 14~TeV High-Luminosity LHC (HL-LHC)~\cite{CERN-LHCC-2015-020}; Appendix~\ref{apndx:generator_studies} shows minor differences between these energies so we generate only 14 TeV MC for simplicity. 

%-------------------------------------
\section{\label{sec:analysis} Analysis proposal}
%-------------------------------------

\begin{figure*}
    \centering
    \includegraphics[width=\textwidth]{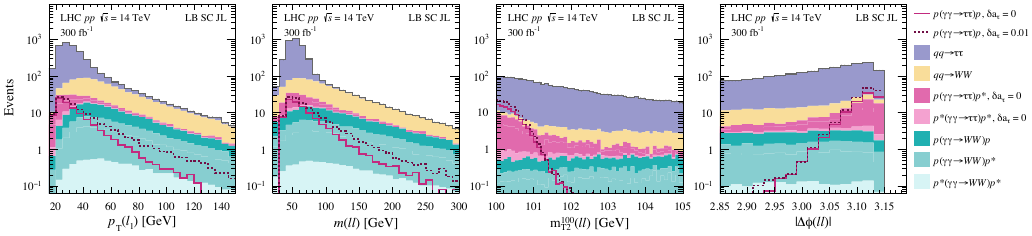}
    \caption{Distributions normalized to 300~fb$^{-1}$ for leading lepton $p_\text{T}^{\ell_1}$ (far left), dilepton mass $m_{\ell\ell}$ (center left), dilepton $m_{T2}^{100}(\ell\ell)$ (centre right) and opening angle $|\Delta \phi_{\ell\ell}|$ (far right). 
    Electron-muon preselection $p_\text{T}^{e\,(\mu)} > 18\,(15)$~GeV, $|\eta_\ell| < 2.5$, and vertex isolation $\Delta z_\text{min} > 1$~mm via $(\varepsilon_\text{PU}, \varepsilon_\text{UE}) = (50\%, 0.4\%)$ are applied.
    SM (BSM) elastic $\gamma\gamma\to\tau\tau$ is the solid (dashed) line; other processes are stacked histograms.}
    \label{fig:emu-kindistro1}
\end{figure*} 

We propose targeting the $\tau\tau \to e\nu\nu \mu\nu\nu$ decay with $2\times \mathcal{B}(\tau \to e \nu \nu) \times \mathcal{B}(\tau \to \mu \nu \nu) \simeq 6\%$~\cite{Tanabashi:2018oca} branching ratio. 
We emulate standard dilepton triggers selecting this signature by requiring the electron (muon) satisfy current offline thresholds for the ATLAS experiment $p_\text{T}^{e\,(\mu)} > 18 \,(15)$~GeV~\cite{ATLAS:2019dpa,ATLAS:2020gty,ATLAS:2021tnq,ATL-DAQ-PUB-2019-001} and tracker acceptance $|\eta_{e, \mu}| < 2.5$.  
After this selection, several backgrounds mimic this signal, which we denote \emph{reducible} (\emph{irreducible}) if they vanish (persist) in the limit of ideal detector efficiency and resolution.
Quark-induced ditau $qq \to \tau\tau \to e\nu\nu\mu\nu\nu$ (Drell-Yan) and diboson $qq \to WW \to e\nu\mu\nu$ dominate reducible backgrounds; photon-induced diboson $\gamma\gamma \to WW \to e\nu\mu\nu$ dominates irreducible backgrounds. 
Reducible backgrounds from non-prompt leptons and misidentified jets are subdominant (below 10\%) in similar analyses~\cite{ATLAS:2020iwi,ATL-PHYS-PUB-2021-026}, whose estimation is beyond the scope of this study focusing on the dominant backgrounds.
We design an (i) \emph{observation analysis} optimized for extracting the SM $\gamma\gamma\to\tau\tau$ signal from backgrounds and (ii) \emph{dipole analysis} targeting BSM deviations amid backgrounds. 

To reject reducible quark-induced backgrounds, we apply the track-vertex isolation technique~\cite{ATLAS:2020mve,ATLAS:2020iwi}.
This requires the electron-muon principal vertex (PV) and track nearest to the PV have longitudinal separation $\Delta z_\text{min} = \left|z_\text{trk}^\text{min} - z_\text{PV}\right|$  (Fig.~\ref{fig:feynGraphs_yy2tautau}, right).
This isolation (track veto) condition selects photon-fusion processes that exhibit no additional tracks associated to the PV, whereas quark-induced backgrounds have many tracks from underlying event (UE) in proton breakup. 
We follow Ref.~\cite{ATL-PHYS-PUB-2021-026} and emulate the $\Delta z_\text{min} > 1~\text{mm}$ condition assuming nominal $p_\text{T}^\text{trk} > 500$~MeV by applying a $\varepsilon_\text{UE} = 0.4\%$ efficiency to our quark-induced MC $(qq\to WW, \tau\tau)$.
Pileup tracks can also accidentally fall within the veto window, so we further apply a $\varepsilon_\text{PU} = 50\%$ efficiency based on data~\cite{ATLAS:2020iwi,ATL-PHYS-PUB-2021-026} to all MC, which depends on the process-independent track density governed by the beamspot width $\sigma_z^\text{beam} \simeq 38$~mm in Run~2~\cite{ATLAS:2020iwi}.
We consider $\varepsilon_\text{UE} = 0.4\%$ as conservative given measurements show this is nearer 0.1\% in $Z\to ee/\mu\mu$ data~\cite{ATLAS:2020iwi}; the difference arises from UE mismodelling in simulation~\cite{ATLAS:2019ocl}. 
We normalize our single (double) dissociative \textsc{MadGraph} MC to 0.5 (0.1) of the elastic cross-section to account for vertex isolation and soft survival based on Ref.~\cite{Harland-Lang:2020veo}.
Reconstructing UE tracks outside nominal acceptance would improve $\varepsilon_\text{UE}$ by (i) increasing tracker acceptance $|\eta^\text{trk}| < 4$ at HL-LHC~\cite{PhaseII:ITkStripsTDR} and (ii) reducing track thresholds to $p_\text{T}^\text{trk} > 200$~MeV. 
A $p_\text{T}^\text{trk} > 100$~MeV threshold is feasible at low luminosity~\cite{ATLAS:2016zba}, but reconstruction at high luminosity practically requires $p_\text{T}^\text{trk} > 200$~MeV to avoid higher fake-track rates if $p_\text{T}^\text{trk}$ were further reduced~\cite{McCormack:2716311}.
Standard data processing requires too much computational resources to run low-$p_\text{T}$ tracking in high-luminosity collisions.
Therefore, low-$p_\text{T}$ tracking requires unconventional data processing that reconstructs a subset of events passing kinematic selection offline rather than all events online, akin to ATLAS \emph{delayed stream} and CMS \emph{data parking} paradigms.
We use a simple $\varepsilon_\text{UE} = 0.04\%$ benchmark to capture these possible improvements.
Increased $p_\text{T}^\text{trk} > 1$~GeV~\cite{ATL-PHYS-PUB-2021-026} and pileup nuance HL-LHC extrapolations beyond the scope of this work, so we capture this simply with $\varepsilon_\text{PU} = 10\%$, following Ref.~\cite{ATL-PHYS-PUB-2021-026}. 
These considerations motivate the $\varepsilon_\text{UE} \in [0.4\%, 0.04\%]$ and $\varepsilon_\text{PU} \in [50\%, 10\%]$ benchmarks in our projections.

Turning to electron-muon kinematics, we henceforth denote $\ell \in [e, \mu]$. 
The electron-muon pair from $\gamma\gamma\to \tau\tau$ is nearly back-to-back, so requiring the azimuthal opening angle satisfy $|\Delta \phi_{\ell\ell}| > 3.1$ provides 77\% signal efficiency with little BSM dependence for 14\% $qq \to \tau\tau$ and 3\% $WW$ efficiency. 
While the dilepton invariant mass $m_{\ell\ell}$ from $qq \to \tau\tau$ peaks around 40~GeV from resonant $Z$-boson production, a looser requirement $m_{\ell\ell} > 20$~GeV retains greater signal efficiency. We also require the stransverse mass defined in Refs.~\cite{Barr:2003rg,Lester:2014yga,ATLAS:2019lng} satisfy $m_\text{T2}^{100} < 101$~GeV. 
The $m_\text{T2}$ variable reconstructs a kinematic endpoint at the mass of the parent state in doubly semi-invisible decays and therefore helps remove $WW \to \ell\nu\ell\nu$ backgrounds versus $\tau\tau \to e\nu\nu \mu\nu\nu$. 
Similar to Ref.~\cite{Beresford:2018pbt}, we use the dilepton $p_\text{T}^{\ell\ell}$ as the missing transverse momentum proxy in $m_\text{T2}^{100}$. Figure~\ref{fig:emu-kindistro1} shows $p_\text{T}^{\ell_1}$, $m_{\ell\ell}$, $m_\text{T2}^{100}$ and $|\Delta \phi_{\ell\ell}|$ distributions motivating the selection. 
To summarize the event selection:
\begin{itemize}
    \item \emph{Observation analysis} considers both elastic and dissociative $\gamma\gamma\to \tau\tau$ with $\delta a_\tau 
    = 0$ as signal following Refs.~\cite{ATLAS:2020iwi,ATL-PHYS-PUB-2021-026}, applying $p_\text{T}^{e\,(\mu)} > 18\,(15)$~GeV, $|\eta_\ell| < 2.5$, $m_{\ell\ell} > 20$~GeV, $\Delta z_\text{min} > 1$~mm (via $\varepsilon_\text{UE}, \varepsilon_\text{PU}$), $|\Delta \phi_{\ell\ell}| > 3.1$, $m_\text{T2}^{100} < 101$~GeV.
    
    \item \emph{Dipole analysis} optimizes for $\delta a_\tau 
    \neq 0$ signals with a simple shape analysis by taking the single-bin cut-and-count \emph{Observation analysis} and bins in leading lepton $p_\text{T}^{e_1\,(\mu_1)} \in [18\,(15), 40, \infty]$ to exploits the harder BSM $p_\text{T}^\ell$ spectrum with a shape analysis (as introduced in Ref.~\cite{Beresford:2019gww}), and subtracts dissociative $\gamma\gamma\to \tau\tau$ to leave only elastic $\gamma\gamma\to \tau\tau$ as signal. 
    We use two bins for simplicity, but future work could improve sensitivity via finer binning. 
\end{itemize}

For systematic uncertainties, we outline strategies for experimentalists to control them with powerful data-driven techniques. 
The dominant background systematic is UE track modelling, where control samples of Drell-Yan $qq \to \ell\ell$ can already constrain $qq \to \tau\tau$ to the level of 5--7\%~\cite{ATLAS:2020iwi}.
The dominant $\gamma\gamma\to\tau\tau$ signal uncertainties arise from initial-state dynamics that we could estimate to be $\mathcal{O}(10\%)$ by comparing photon fluxes (Appendix~\ref{apndx:systematics}) or alternative generators~\cite{Harland-Lang:2018iur,Harland-Lang:2020veo,Harland-Lang:2019pla}.
Fortunately, we can constrain these by measuring the high-statistics standard-candle $\gamma\gamma\to \ell\ell$ process (1.1~pb with $p_\text{T}(e/\mu) > 18/15$~GeV, $3.3 \times 10^5$ events at 300~fb$^{-1}$) to correct both absolute $\gamma\gamma\to \tau\tau$ cross-sections and generator-level kinematics important for dipole shape analyses.
Deriving such corrections from $\gamma\gamma\to ee/\mu\mu$ for $\gamma\gamma\to\tau\tau$ is well justified given the flavor independence of both initial-state proton soft-survival physics and QED final-state radiation. 
ATLAS has deployed these control techniques in similar measurements~\cite{ATLAS:2022ryk,ATLAS:2020iwi}. We emphasize measuring dilepton mass $m_{\ell\ell}$ and rapidity $y_{\ell\ell}$ in distinct azimuthal angle $|\Delta \phi_{\ell\ell}|$ regions (or the correlated $p_\text{T}^{\ell\ell}$) would probe the composition of dissociative $\text{pp} \to \text{p}(\gamma\gamma\to \ell\ell)\text{p}^*$ enhanced at low $|\Delta \phi_{\ell\ell}| < 3.1$.
This ensures simulation can accurately model and subtract dissociative production so only the elastic $\gamma\gamma\to\tau\tau$ signal remains for measuring EM dipoles in the $q^2_\gamma \to 0$ limit.
From recent $\gamma\gamma\to \ell\ell$ measurements~\cite{Aaboud:2017oiq,ATLAS:2020mve}, we anticipate such data-driven corrections to $\gamma\gamma\to \tau\tau$ predictions can reach percent-level accuracy.
With HL-LHC statistics, we anticipate experimental systematics to be limiting: luminosity uncertainties already reach 0.8\% precision~\cite{ATLAS:2022hro}, while electron/muon momentum calibration will limit kinematic corrections, where the $ee$ channel and upgraded trackers will enable tuning in $2.5<|\eta_e|<4$.
Together, this motivates the assumed [1\% 5\%] systematics for benchmarking our projections, similar to the approach of Ref.~\cite{ATL-PHYS-PUB-2021-026}.

%-------------------------------------
\section{\label{sec:results} Sensitivity projections}
%-------------------------------------

Applying the event selection and assuming 300~fb$^{-1}$ luminosity, we find our \emph{observation analysis} yields $S=120$ signal and $B=440$ background counts ($S/B = 0.27$) assuming $(\varepsilon_\text{UE}, \varepsilon_\text{PU}) = (0.4\%, 50\%)$.
Quark-induced processes $qq \to\tau\tau$ (96\%, 425 events) and $qq \to WW$ (3\%, 13 events) dominate the background composition with $\gamma\gamma\to WW$ in remainder.
Evaluating the Poissonian asymptotic significance $Z_A(S, B, \zeta_b)$~\cite{Cowan:poissonunc,Cowan:2010js} assuming $\zeta_b$ background systematics, this exceeds five standard deviations $Z_A = 5.4$ for $\zeta_b = 1\%$.
Assuming the more accurate value of $\varepsilon_\text{UE} = 0.1\%$, $Z_A = 8.5$ is feasible for $\zeta_b = 5\%$.
This establishes a procedure to observe $\gamma\gamma\to \tau\tau$ using standard pp runs and tracking.

\begin{figure}
    \centering
    \includegraphics[width=0.49\textwidth]{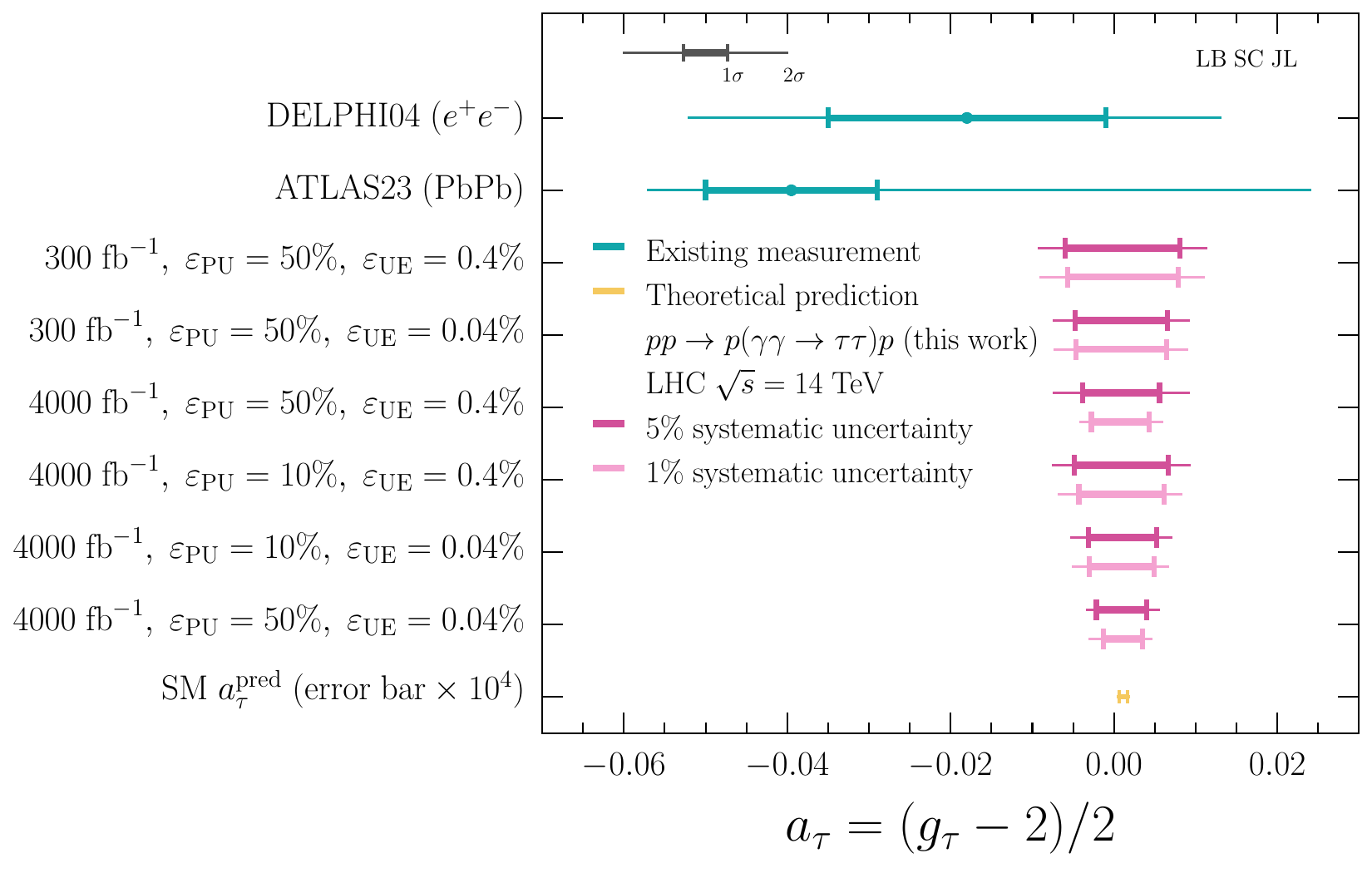}
    \caption{Summary of tau-lepton magnetic dipole $a_\tau = (g_\tau -2)/2$ sensitivity. 
    Our projections (pink) assume 300~fb$^{-1}$ and 4000~fb$^{-1}$ luminosities for various pileup $\varepsilon_\text{PU}$ and underlying-event $\varepsilon_\text{UE}$ efficiencies with 5\% (dark) and 1\% (light) systematics. 
    Displayed are existing DELPHI~\cite{Abdallah:2003xd} and ATLAS~\cite{ATLAS:2022ryk} constraints (blue) alongside the SM prediction $a_\tau^\text{pred}$ (orange)~\cite{Eidelman:2007sb}.
    The thick (thin) lines indicate 68\% CL (95\% CL) limits.}
    \label{fig:atau_limits_summary}
\end{figure}

We estimate our \emph{dipole analysis} sensitivity to BSM $\delta a_\tau$ using a $\chi^2$ statistical test against the SM null hypothesis 
\begin{align}
    \chi^2_r = \frac{(S_\text{SM+BSM} - S_\text{SM})^2}{\sigma_\text{stat}^2 + \sigma_\text{syst}^2}.
\end{align}
We evaluate region $r$ with $B$ background yield, $S_\text{SM}$ ($S_\text{SM+BSM}$) signal yield assuming SM (nonzero $\delta a_\tau$) couplings, statistical $\sigma_\text{stat}^2 = B + S_\text{SM+BSM}$ and systematic $ \sigma_\text{syst}^2 = (\zeta_b B)^2 + (\zeta_s S_\text{SM+BSM})^2$ uncertainties parameterized by $\zeta_{b,s}$.
We consider a simplified scheme that assumes $\zeta = \zeta_s = \zeta_b$ for the benchmarks $\zeta \in [1\%, 5\%]$ and are uncorrelated to statistically combine using $\chi^2 = \sum_\text{r} \chi_\text{r}^2$. 
The inequality $\chi^2 < 1$ ($\chi^2 < 3.84$) defines the 68\% CL (95\% CL) constraints on $\delta a_\tau$.
Our results focus on $a_\tau$ where the LHC is competitive, whereas Belle remains most competitive for $d_\tau$~\cite{Beresford:2019gww,Inami:2002ah}. 

Figure~\ref{fig:atau_limits_summary} summarizes our projected sensitivity for $a_\tau = a_\tau^\text{SM} + \delta a_\tau$. 
With 300~fb$^{-1}$ luminosity, we find $-0.0092 < a_\tau < 0.011$ at 95\% CL for the benchmarks $(\zeta, \varepsilon_\text{PU}, \varepsilon_\text{UE})= (5\%, 50\%, 0.4\%)$.
This is a fourfold improvement in magnitude over existing LEP~\cite{Abdallah:2003xd,1998169,Ackerstaff:1998mt} and LHC~\cite{ATLAS:2022ryk,CMS:2022arf} limits.
Setting $C_{\tau B} = -1$ as a benchmark in Eq.~\eqref{eq:delta_a_d_tau_defn}, this corresponds to new physics scale sensitivity reaching $\Lambda^{95}_{C=-1} > 430$~GeV at 95\% CL (from $\Lambda^{95}_{C=-1} > 190$~GeV at LEP).
For 4000~fb$^{-1}$ HL-LHC extrapolation, we also loosen $p_\text{T}^\ell > 10$~GeV, $|\eta_e| < 4$~GeV~\cite{PhaseII:TDAQTDR} to find an eightfold improvement beyond LEP $-0.0029 < a_\tau < 0.0046$ for the most optimistic scenario $(\zeta, \varepsilon_\text{PU}, \varepsilon_\text{UE})= (1\%, 50\%, 0.04\%)$, corresponding to $\Lambda^{95}_{C=-1} > 680$~GeV.
This HL-LHC reach degrades modestly to $-0.0067 < a_\tau < 0.0082$ with more realistic benchmarks $(\zeta, \varepsilon_\text{PU}, \varepsilon_\text{UE})= (1\%, 10\%, 0.4\%)$.
The SMEFT prescription used for such benchmark interpretations is only valid at probed energy scales below $\Lambda$ to satisfy unitarity; given the weak experimental bounds on $(a_\tau, d_\tau)$, this implies strong couplings $C_{\tau B} \sim \mathcal{O}(1)$.
It remains an interesting model-building challenge for future work in construct concrete BSM theories that match to these SMEFT operators with such large couplings.
Our results nonetheless establish an important strategy to overcome the longstanding obstacles of measuring $\gamma\gamma\to\tau\tau$ and $a_\tau$ in standard LHC runs.
This introduces a novel avenue toward BSM physics via precision competitive with the landmark one-loop QED prediction $\alpha_\text{EM}/2\pi \simeq 0.0012$. 

%-------------------------------------
%\section{\label{sec:summary}Summary}
%------------------------------------- 
In summary, we proposed the strategy to measure $\gamma\gamma\to \tau\tau$ in LHC proton beams, with  $-0.0092 < a_\tau < 0.011$ (95\% CL) dipole sensitivity assuming 300~fb$^{-1}$ luminosity and 5\% systematics.
This opens future work to develop dedicated triggers strategies  with reduced $p_\text{T}$ thresholds to increase photon-induced ditau yields, machine learning such as multi-class graph neural networks to improve tau-lepton identification with reduced jet misidentification rates, CP-sensitive observables for electric dipoles, and combinations with ALICE, LHCb, and heavy ions.
Experimental realization would furnish a novel precision tau-lepton dipole program that could reveal new physics in quantum fluctuations. 

\emph{Note added}: a strategy developed independently by the CMS Collaboration to measure $\gamma\gamma\to \tau\tau$ in pp data appeared soon after our paper was released~\cite{CMS:2024skm}. 

%-------------------------------------
\begin{acknowledgements}
%-------------------------------------
%\vspace{-2ex}
%\emph{\textbf{Acknowledgements}}---
We thank Will Barter, Markus Diehl, Aleksandra Dimitrievska, Mateusz Dyndal, Zahra Ghorbanimoghaddam, Lucian Harland-Lang, Oldrich Kepka, Jakub Kremer, Valerie Lang, Kristin Lohwasser, Maeve Madigan, Klaus M\"{o}nig, Anna Mullin, Simone Pagan Griso, Andrew Pilkington, Philip Sommer, and Weronika Stanek-Maslouska for helpful discussions. 
We are grateful to the hospitality of the Institute for Particle Physics Phenomenology at Durham University hosting the \emph{Photon-induced Processes Workshop}, and University of Hamburg hosting the \emph{European Physical Society Conference on High Energy Physics}, which facilitated in-person collaboration. 
LB and SC are supported by the Helmholtz Association ``Young Investigator Group'' initiative. 
JL is supported by a Junior Research Fellowship at Trinity College, University of Cambridge. 
\end{acknowledgements}

%\onecolumngrid

%-------------------------------------
\bibliography{bibs/intro.bib,bibs/pheno.bib,bibs/software.bib,./bibs/exp.bib,./bibs/theory.bib,./bibs/upc.bib}
%-------------------------------------

\appendix
\clearpage
\section*{Supplementary Material}

This Supplementary Material provides supplementary details supporting the main text.
Section~\ref{apndx:MCdetails} provides further methodological details on the simulation of signal and background processes.
Section~\ref{apndx:systematics} expands the discussion on theoretical systematic uncertainties.
Section~\ref{apndx:generator_studies} shows validation studies for generator-level distributions.

%-------------------------------------------------
\subsection{\label{apndx:MCdetails}Technical simulation details}
%-------------------------------------------------

For the elastic $\gamma\gamma\to\tau\tau$ process, we generate around 2 million MC events for each coupling variation and require $p_\text{T}^{\tau} > 15$ GeV in \textsc{MadGraph} by setting \texttt{\{15:15\}=pt\_min\_pdg} in the \texttt{run\_card} to improve generator statistics.
The SM cross-section reduces from 150~pb to 0.72~pb after imposing $p_\text{T}^{\tau} > 15$ GeV, corresponding to an efficiency of 0.5\%.
We further improve generator statistics by requiring tau-leptons decay fully leptonically in \textsc{Pythia}~8.306~\cite{Sjostrand:2007gs,Bierlich:2022pfr}  using:\\ 
\texttt{15:onMode = off}\\
\texttt{15:onIfAny = 11 13}.\\
We then account for the dileptonic branching fraction $\mathcal{B}(\tau\tau \to \ell\nu\nu\ell\nu\nu) \simeq 12.4\%$ in cross-section normalization. 
For elastic processes, the physical picture of the $q_\gamma \to 0$ limit is that the EM fields surrounding the protons not only comprise the photons pair creating tau-leptons but also provide the quasi-static external EM field in which we measure the tau-lepton EM dipoles.

Single dissociation (elastic-inelastic photon fusion $\text{pp} \to \text{p}(\gamma\gamma \to \tau\tau)\text{p}^*$ where one proton dissociates) and double dissociation (inelastic-inelastic photon fusion $\text{pp} \to \text{p}^*(\gamma\gamma \to \tau\tau)\text{p}^*$ where both protons dissociate) processes are not implemented in the \textsc{gammaUPC}~1.0 package inside \textsc{MadGraph}~3.5.0.
Therefore, we use \textsc{MadGraph}~2.6.7 based on Ref.~\cite{ATLAS:2020iwi} to implement the equivalent photon approximation~\cite{Budnev:1974de} prescription for the elastic photon $\gamma$, and the default \textsc{NNPDF23\_lo\_as\_0130\_qed} parton distribution functions (PDF)~\cite{Ball:2012cx} for the inelastic photon $\gamma^*$.
We find the \textsc{MadGraph} cross-section for single (double) dissociative photon-fusion production of tau pairs to be 3.41~pb (2.58~pb) with a $p_\text{T}^{\tau} > 15$~GeV generator requirement. 
For the $\gamma\gamma \to WW$ process, no generator cuts are imposed and we find the elastic (single dissociative) cross-section to be 74.1~fb (592~fb).
To decay, shower and hadronize such processes, we use \textsc{Pythia} configured specifically for photon-fusion processes generated in \textsc{MadGraph}. 
The full \textsc{Pythia} settings follow Refs.~\cite{ippp-workshop2023,ATLAS:2020iwi}: 
\begin{itemize}
    \item \texttt{PartonLevel:ISR = off} for elastic processes;
    \item \texttt{BeamRemnants:unresolvedHadron} set to 0, 1/2, 3 for elastic, forward/backward single dissociation and double dissociation, respectively;
    \item All dissociation samples apply these settings:\\
\texttt{PartonLevel:ISR = on}\\
\texttt{BeamRemnants:primordialKThard    = 1.5}\\
\texttt{BeamRemnants:primordialKTremnant = 0.1}\\
\texttt{SpaceShower:dipoleRecoil = on}\\
\texttt{SpaceShower:pTmaxMatch   = 0}\\
\texttt{SpaceShower:pTmaxFudge   = 1.0}\\
\texttt{SpaceShower:pTdampMatch  = 1}\\
\texttt{SpaceShower:pTdampFudge  = 1.5}\\
\end{itemize}

For other non-photon-induced background processes, we generate the Drell-Yan $qq \to Z^{(*)}/\gamma^* \to \tau\tau$ (which we also denote $qq \to \tau\tau$ in the main text) and diboson $qq \to WW \to \ell\nu\ell\nu$ processes to leading order using \textsc{MadGraph}~3.5.0. 
We employ the default NN23LO PDFs~\cite{Ball:2012cx} with up to one parton in the matrix element interfaced to \textsc{Pythia} for parton shower and hadronization using MLM jet-parton matching~\cite{Mangano:2006rw} with merging scale \texttt{xqcut} set to 25~GeV. 
To improve generator statistics, we decay the tau-leptons leptonically $\tau \to \ell \nu\nu$ for $\ell \in [e, \mu]$ and impose a generator-level $p_\text{T}^\ell > 5$~GeV cut in \textsc{MadGraph} (\texttt{5=ptl} in \texttt{run\_card}).
After such requirements, the cross-section for these Drell-Yan (diboson) samples is 154~pb (12.6~pb).
Top quark pair production is observed to be negligible in $e\mu$ selections after track-vertex isolation~\cite{ATLAS:2020mve,ATLAS:2020iwi} given the high track multiplicities of the two heavy-flavor jets, so we avoid simulating this. 

%-------------------------------------------------
\subsection{\label{apndx:systematics}Theoretical systematic uncertainties} 
%-------------------------------------------------

For the $\gamma\gamma\to\tau\tau$ signal, the elementary cross-section is crucially well predicted by quantum electrodynamics and we expect the dominant modeling uncertainties arise from the photon flux and rescattering effects involving the outgoing protons called soft survival. 
Uncertainties in the \textsc{gamma-UPC} photon flux can be theoretically estimated using alternative form factors such as the Electric Dipole (\texttt{EDFF}) model studied in Ref.~\cite{Shao:2022cly}, which shows our baseline (\texttt{ChFF}) more accurately models the data but has a slightly harder $p_\text{T}^\tau$ and $m_{\tau\tau}$ spectrum than the other photon-flux models.
For $\text{pp} \to \text{p}(\gamma\gamma\to\tau\tau)\text{p}$ with $p_\text{T}^\tau > 15~\text{GeV}$, we find using
\texttt{EDFF} gives a cross-section of 0.594~pb that is 18\% smaller than our \texttt{ChFF} baseline of 0.720~pb.
We also compare these kinematic differences in Fig.~\ref{fig:UnitNorm_LHE_yyflux}, which provides a quantitative indication of the theoretical modeling in state-of-the-art generators. 
We find these discrepancies are within 10\% for $m_{\tau\tau} \lesssim 200$~GeV and $p_\text{T}^\tau \lesssim 100$~GeV, but grow to higher values reaching over 30\%. 
Further estimation of theoretical uncertainties could be made by comparing with alternative generators such as \textsc{SuperChic}~\cite{Harland-Lang:2020veo}.
Dissociative processes additionally have uncertainties arising from the inelastic photon PDF, where we could use alternative PDF sets such as \textsc{MMHT2015qed\_nlo}~\cite{Harland-Lang:2019pla}.
Precise evaluation of these systematic uncertainties is beyond the scope of this phenomenological study.
The main text instead discusses in detail the strategies to experimentally constrain these in situ with data to what we expect to reach percent-level accuracy.

%-------------------------------------------------
\subsection{\label{apndx:generator_studies}Generator-level validation studies}
%-------------------------------------------------

This section provides further kinematic validation of the simulated Monte Carlo samples, specifically the generator-level (also called ``truth-level'' in the literature) distributions, inspecting the Les Houches Events (LHE) files using \textsc{pylhe}~\cite{lukas_2018}.

\begin{figure*}
    \centering
    \includegraphics[width=0.33\textwidth]{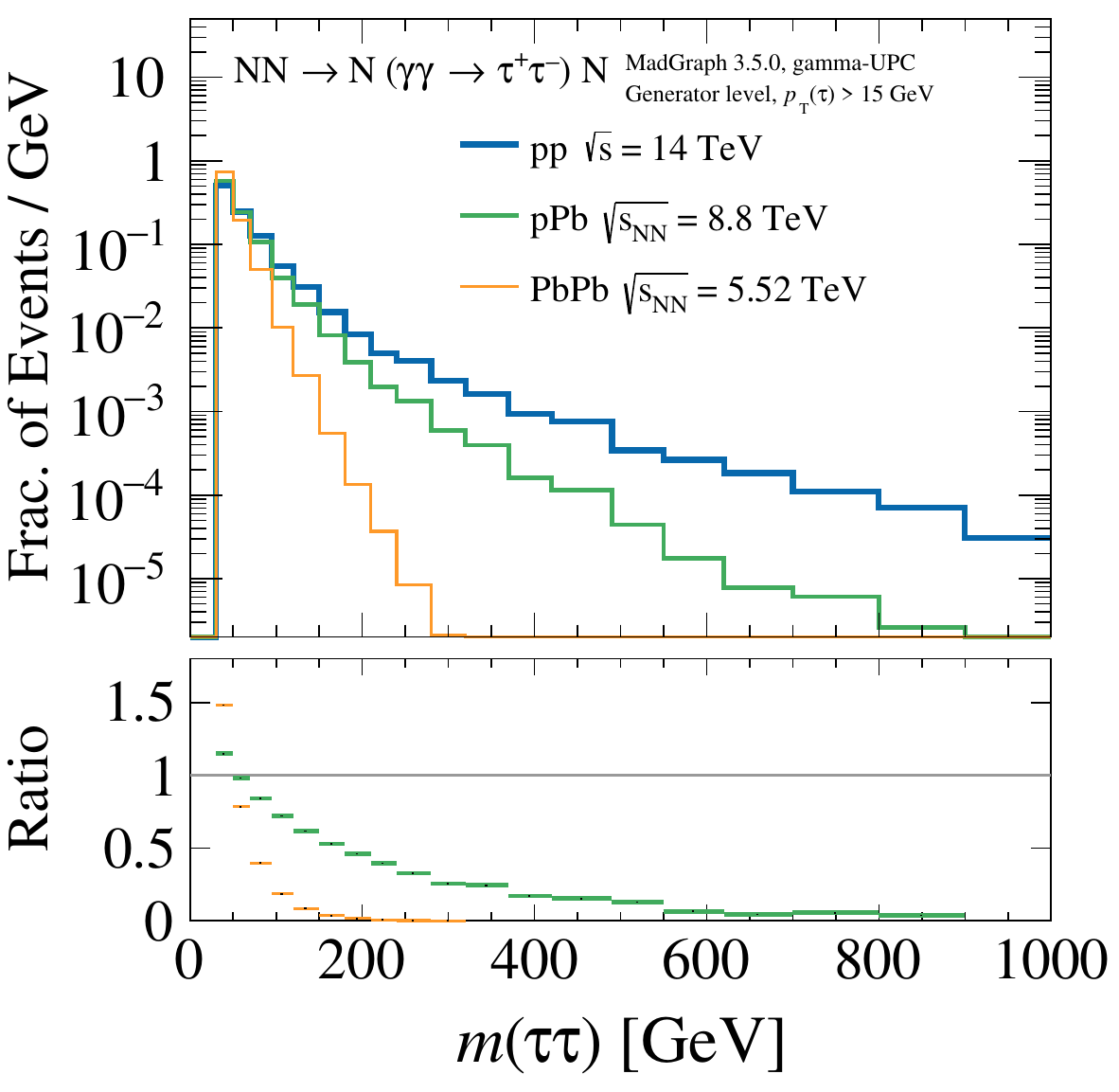}%
    \includegraphics[width=0.33\textwidth]{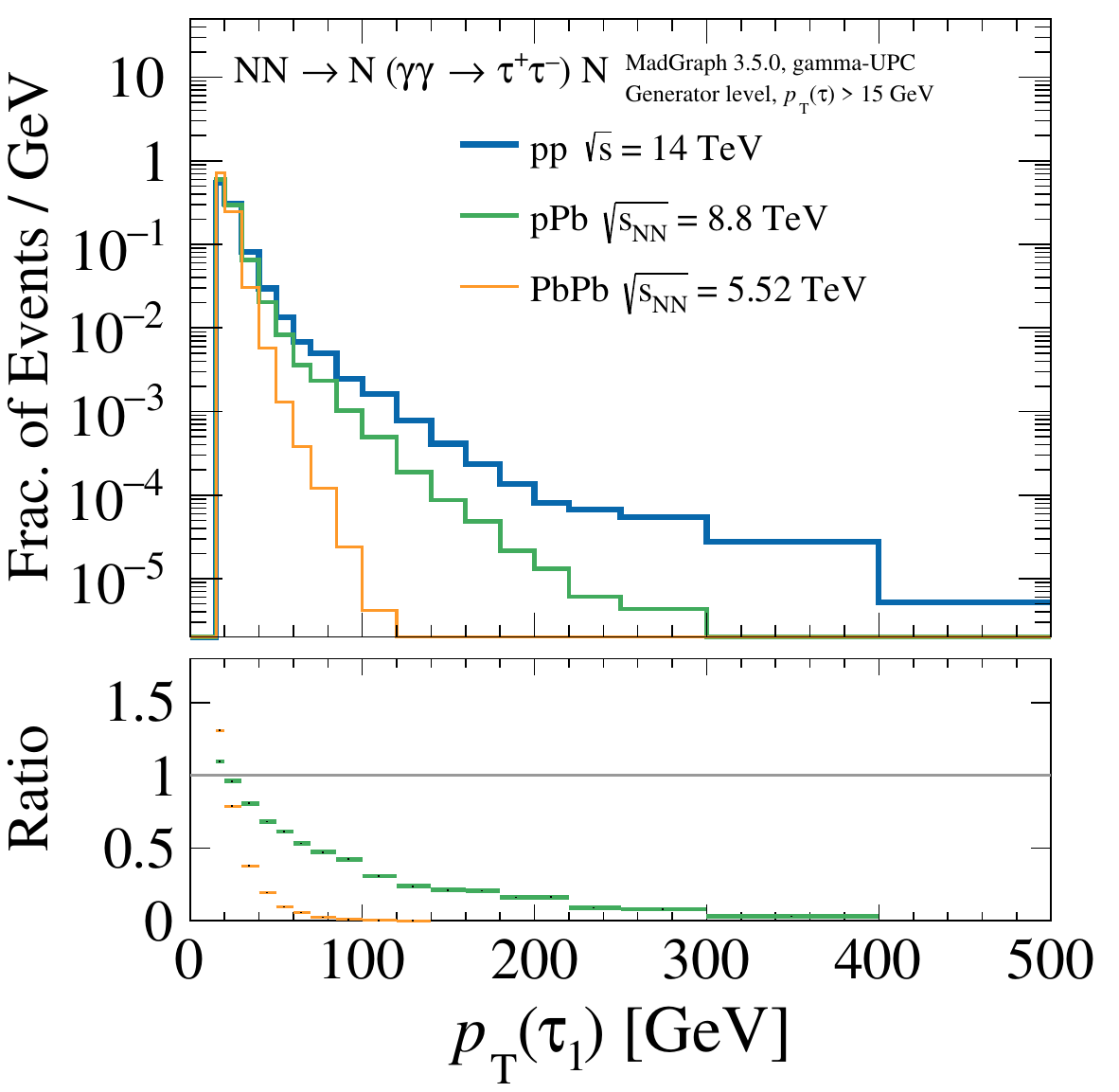}%
    \includegraphics[width=0.33\textwidth]{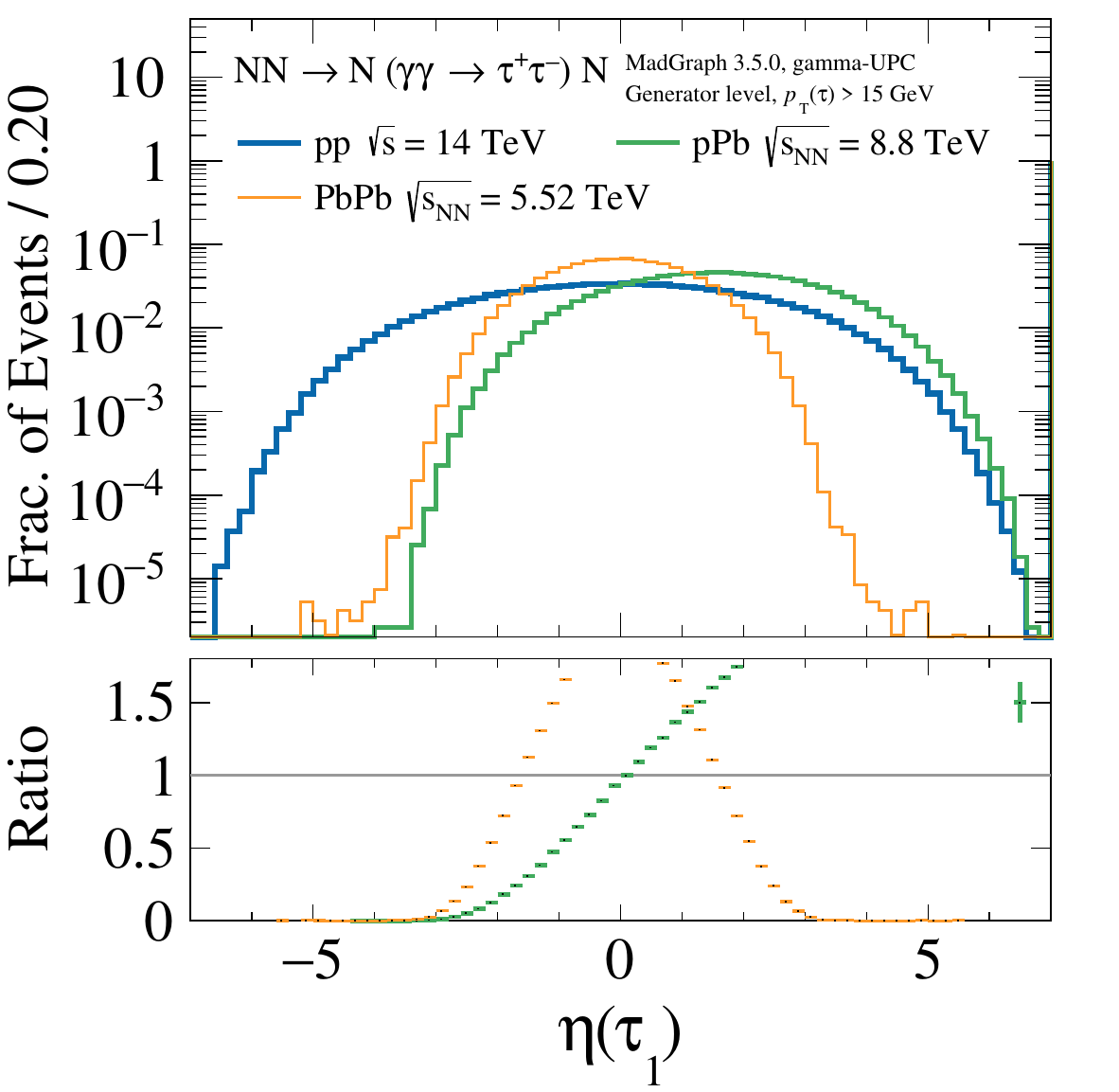}
    \caption{Unit normalized generator-level \textsc{MadGraph} distributions of $\gamma\gamma\to\tau\tau$ comparing beam types: $pp$ (thin blue), proton-lead pPb (medium green), lead-lead PbPb (thick orange). 
    A generator-level requirement of $p_\text{T}(\tau) > 15$ GeV is imposed using the charged form-factor photon flux from \textsc{gamma-UPC}. 
    The lower panel shows the ratio relative to the nominal $pp$.}
    \label{fig:UnitNorm_LHE_beamtype}
\end{figure*}

To highlight the principal advantage of proton-proton over lead-ion collisions, Fig.~\ref{fig:UnitNorm_LHE_beamtype} illustrates kinematic distributions of the $NN \to N(\gamma\gamma\to \tau\tau)N$ process for three beam types $N$ alongside their corresponding center-of-momentum energies (per nucleon for ions): proton-proton pp (14 TeV), proton-lead pPb (8.8 TeV), lead-lead PbPb (5.52 TeV).
We normalize the distributions to unity to compare shapes for the ditau invariant mass $m_{\tau\tau}$, tau-lepton transverse momentum $p_\text{T}^{\tau}$ and rapidity $\eta_{\tau}$.
These variables follow the photon flux distributions, where we consider the charged form factor (ChFF) from \textsc{gamma-UPC} in \textsc{MadGraph}.
The pp collisions reach far higher values in $m_{\tau\tau}$ and $p_\text{T}^{\tau}$ compared those with lead ions.
Relative to $m_{\tau\tau} = 30$~GeV, the $m_{\tau\tau}$ spectrum drops by four orders of magnitude at 200~GeV, 400~GeV and 1000~GeV for PbPb, pPb and pp, respectively.
Similarly, the $p_\text{T}^{\tau}$ spectrum for pp is significantly harder compared to PbPb, where the $p_\text{T}^{\tau}$ can probe above 200~GeV for pp that is inaccessible to PbPb. 
Moreover, the tau-leptons in elastic production are back-to-back in the transverse plane with little transverse boost of the ditau system, so the kinematic boost of the individual tau-leptons arises from the center-of-mass system of the initial-state photons.
The $\eta_{\tau}$ distribution highlights how the larger beam energy of pp gives a higher longitudinal boost than PbPb, while the pPb verifies the expected asymmetric distribution, which illustrate suitability for the LHCb experimental acceptance $2 < |\eta| < 5$ and upgraded trackers for ATLAS and CMS covering $|\eta| < 4$. 

\begin{figure*}
    \centering
    \includegraphics[width=0.33\textwidth]{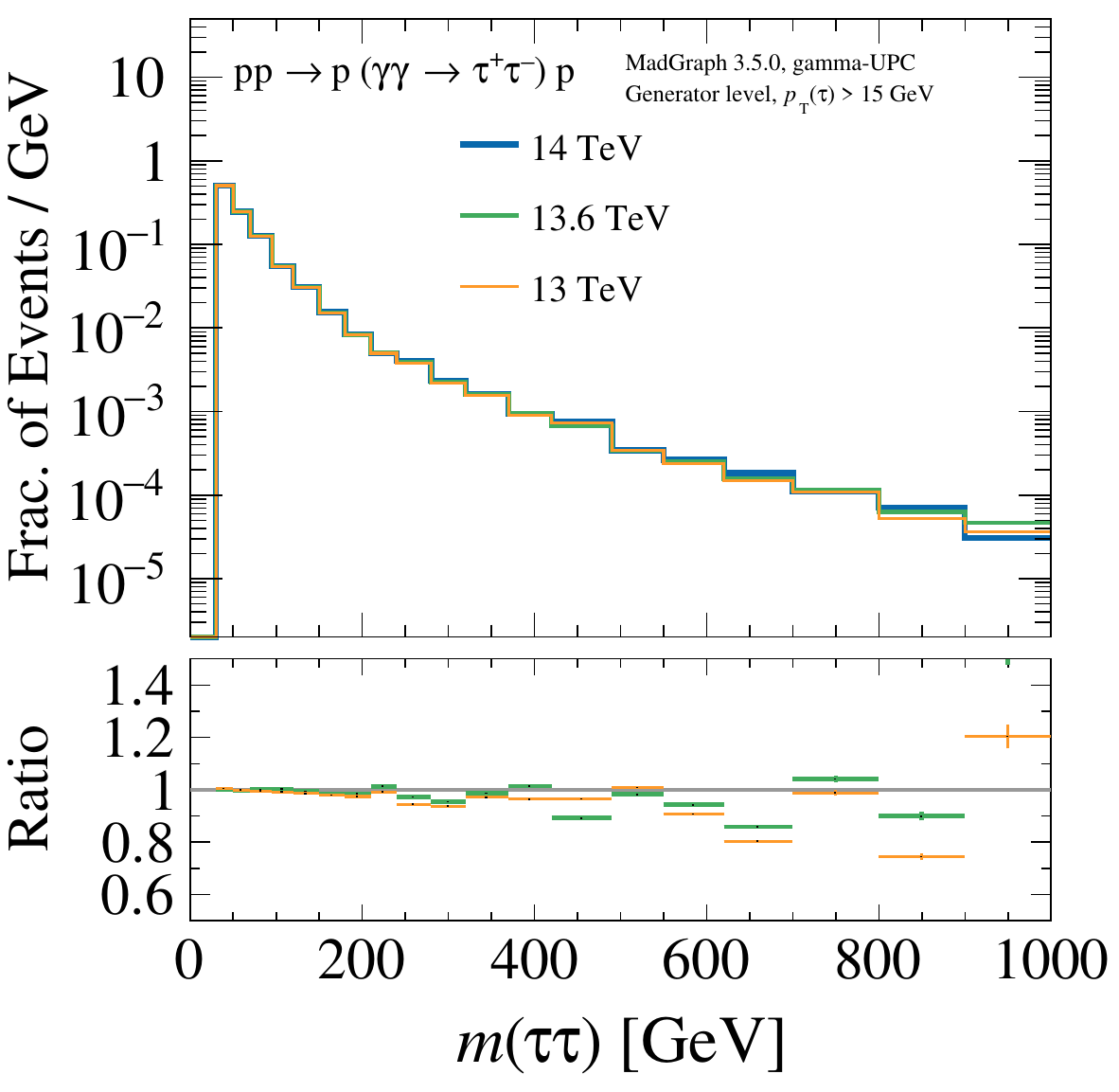}%
    \includegraphics[width=0.33\textwidth]{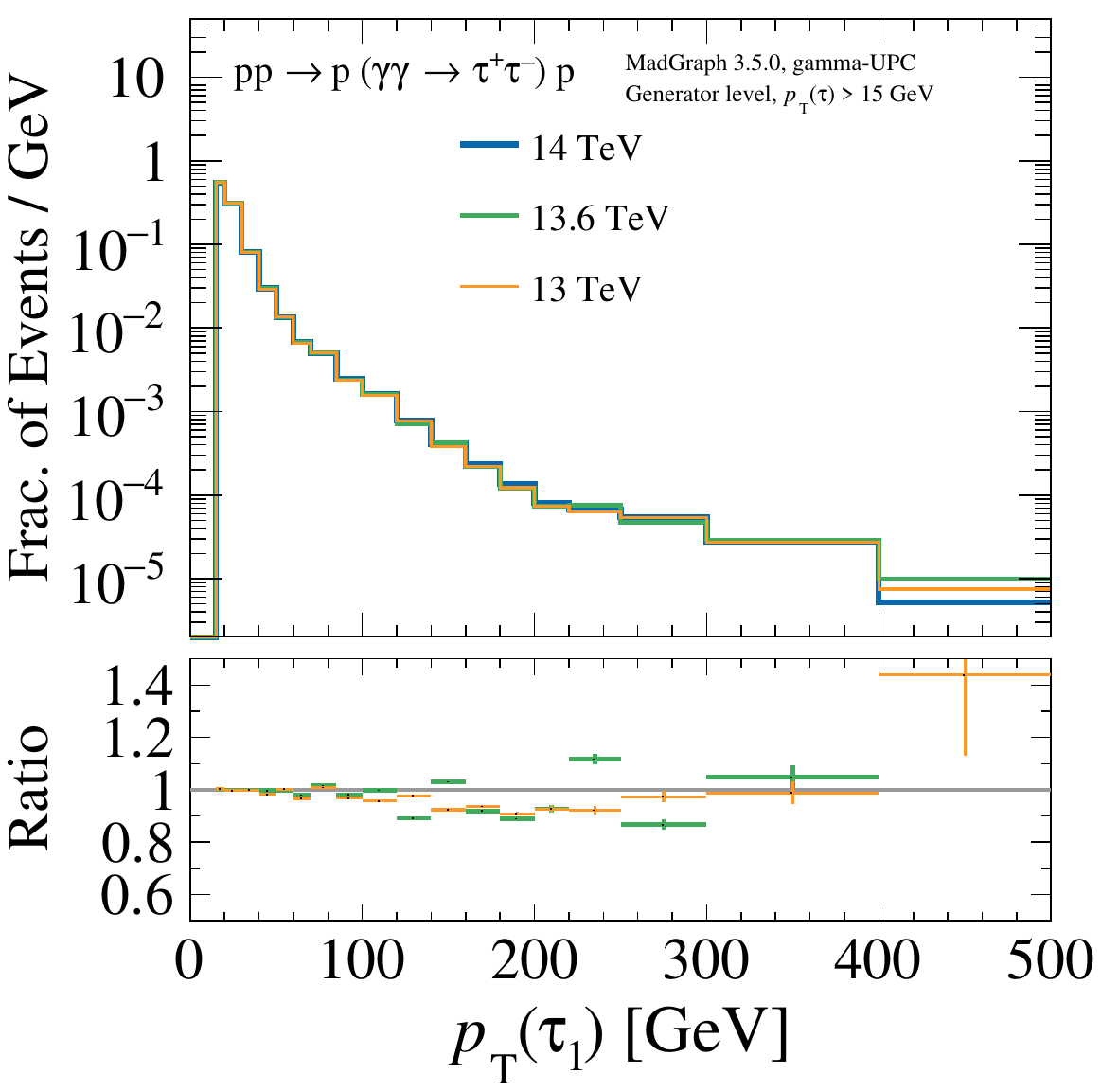}%
    \includegraphics[width=0.33\textwidth]{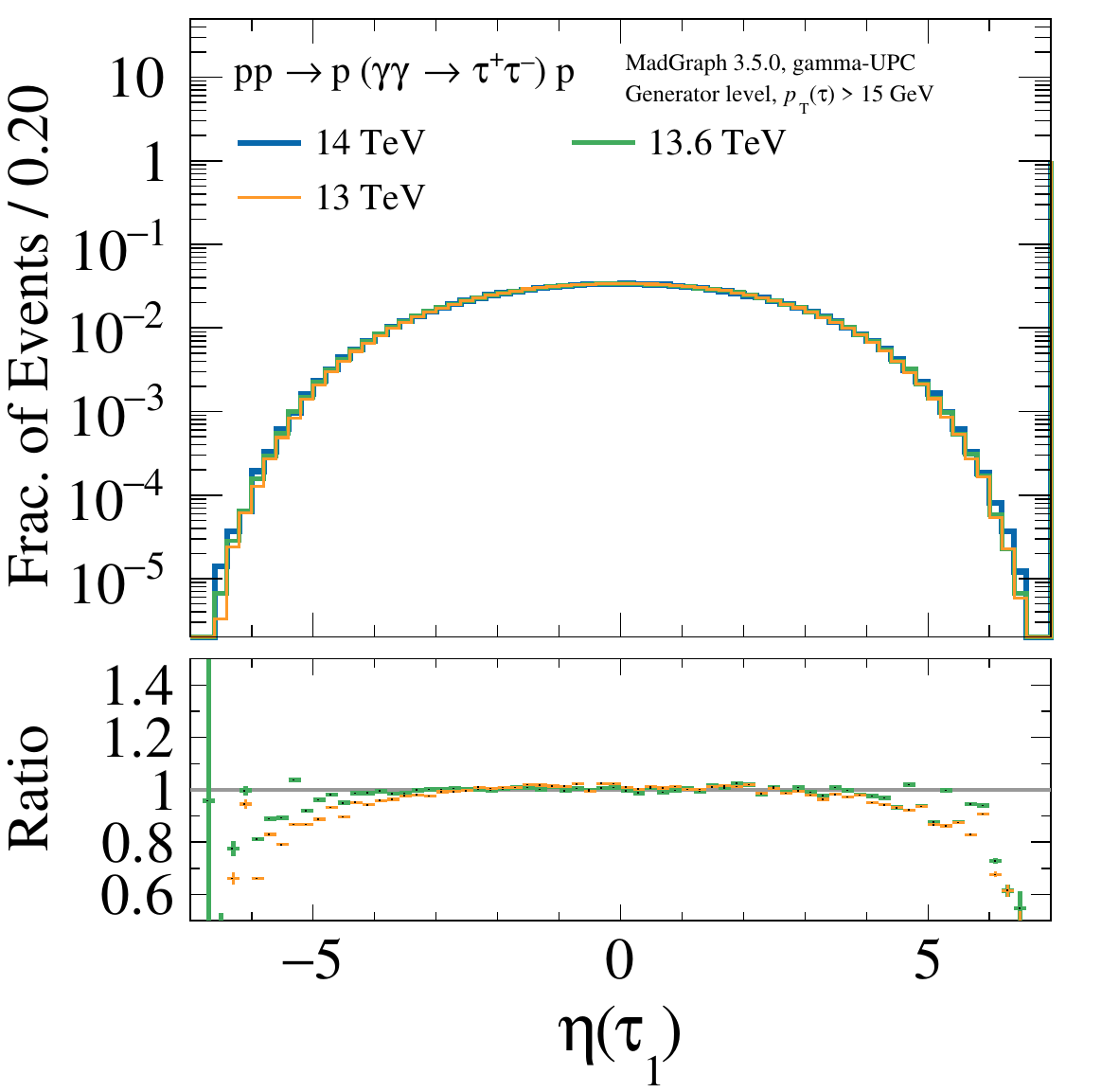}
    \caption{Unit normalized generator-level \textsc{MadGraph} distributions of $\gamma\gamma\to\tau\tau$ comparing $pp$ centre-of-mass energy: 14 TeV (thick blue), 13.6 TeV (medium green), 13 TeV (thin orange). 
    A generator-level requirement of $p_\text{T}(\tau) > 15$ GeV is imposed using the charged form-factor photon flux from \textsc{gamma-UPC}. 
    The lower panel shows the ratio relative to the nominal 14 TeV.}
    \label{fig:UnitNorm_LHE_ppsqrts}
\end{figure*}

We study the impact of the different LHC pp center-of-mass energies $\sqrt{s} \in \{13, 13.6, 14\}$~TeV on distributions of the $\text{pp} \to \text{p}(\gamma\gamma\to \tau\tau)\text{p}$ process in Fig.~\ref{fig:UnitNorm_LHE_ppsqrts}.
This illustrates the differences in $\sqrt{s}$ are generally small within the scope of this study.
Therefore for simplicity, we generate samples at only 14~TeV and rescale luminosities to estimate LHC and HL-LHC sensitivity in the main text.
This is only relevant in the short term where we anticipate the ATLAS and CMS collaborations would combine the 13 TeV from Run 2 with the 13.6~TeV dataset from Run 3, while HL-LHC is expected to reach 14 TeV. 
Crucially, the slight suppression in rate for 13~TeV compared to 14~TeV at the highest $m_{\tau\tau}$ bins are around $\mathcal{O}(10\%)$, which remain subdominant compared to the suppression when using lead ion fluxes (Fig.~\ref{fig:UnitNorm_LHE_beamtype}). 
The suppression at the highest $|\eta_{\tau}|$ values is expected for the reduced longitudinal boost of smaller beam energies, but are outside the close to negligible for the $|\eta| < 4$ acceptance of the upgraded ATLAS and CMS tracker. 

\begin{figure*}
    \centering
    \includegraphics[width=0.33\textwidth]{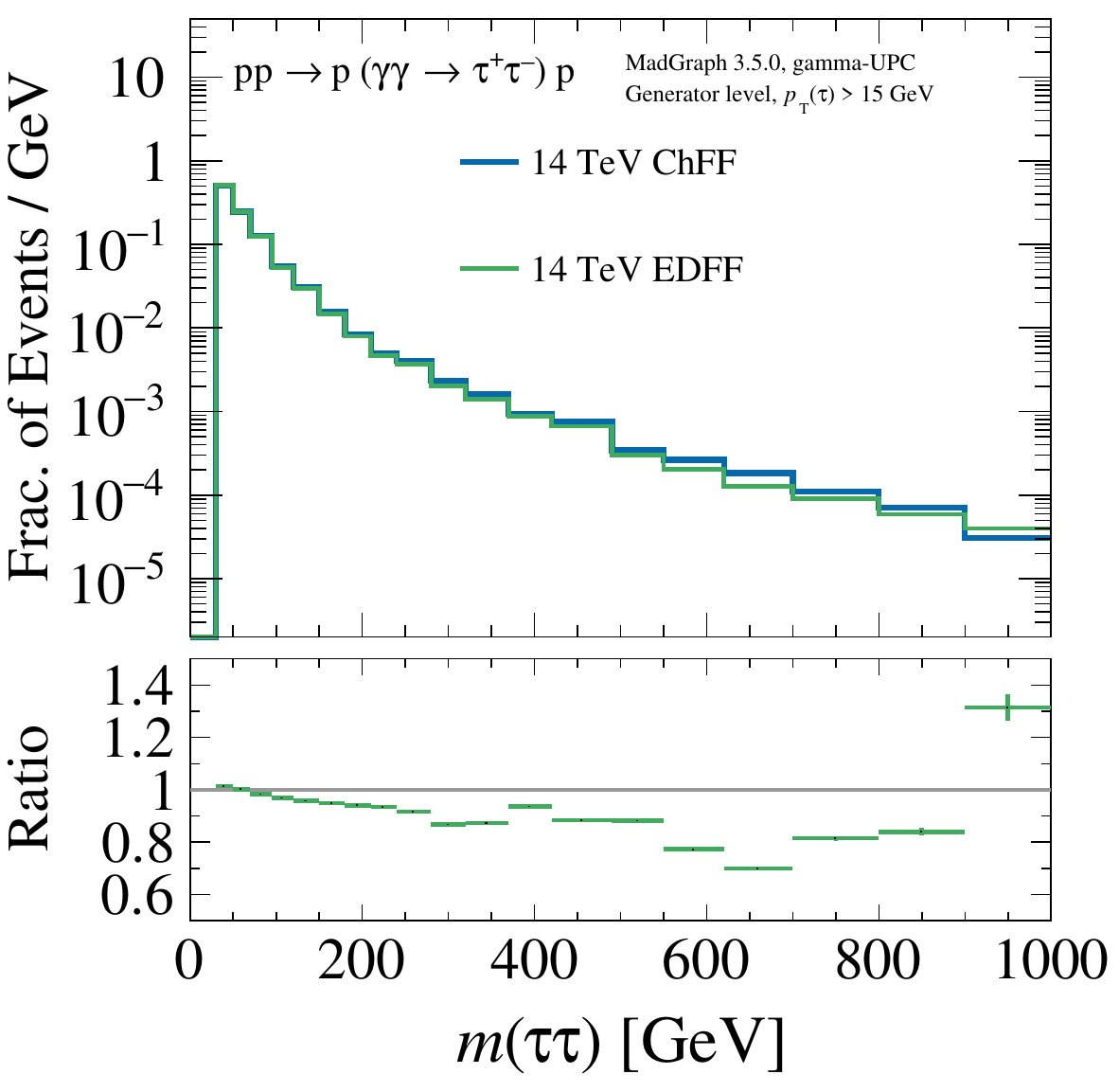}%
    \includegraphics[width=0.33\textwidth]{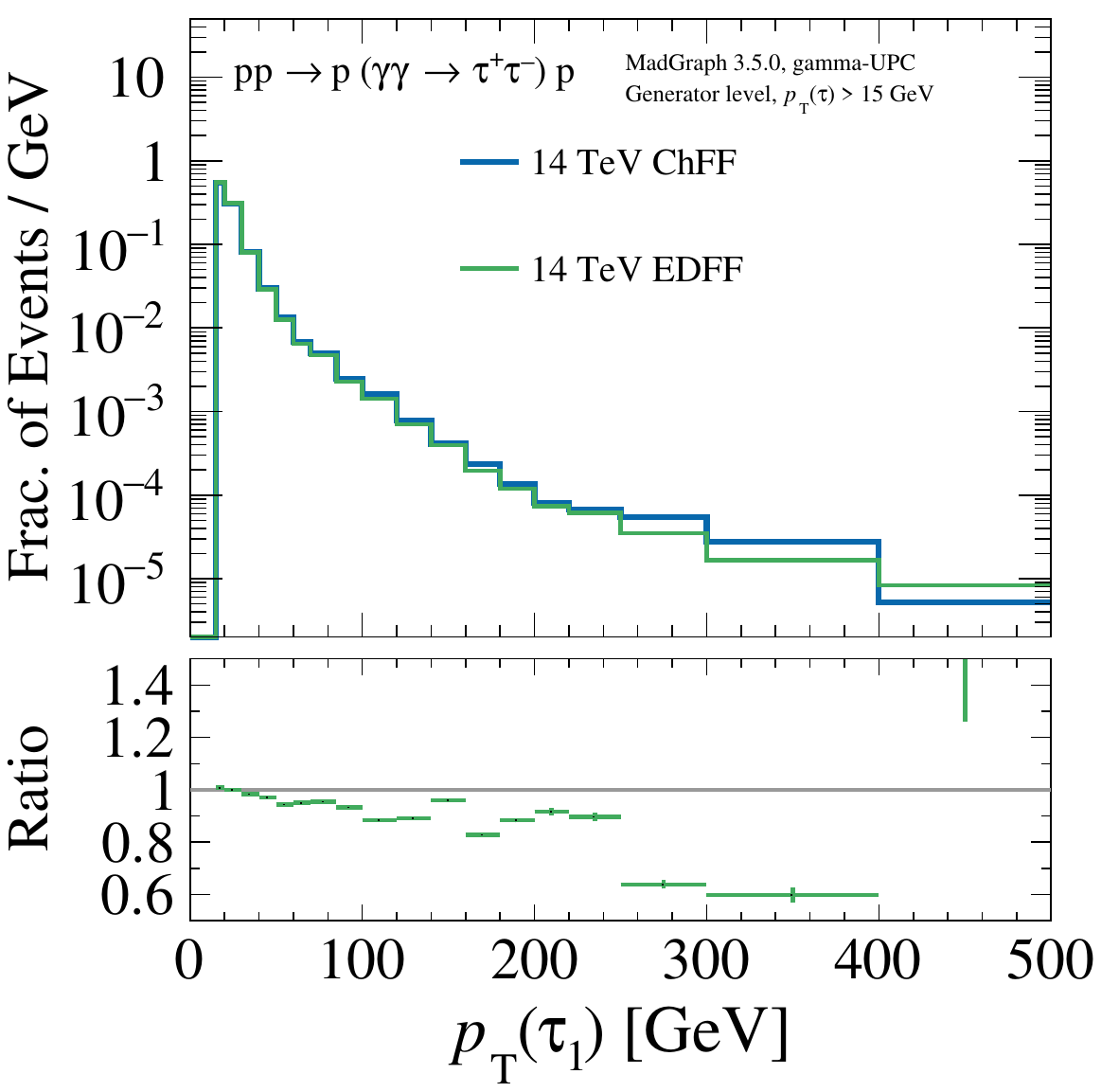}%
    \includegraphics[width=0.33\textwidth]{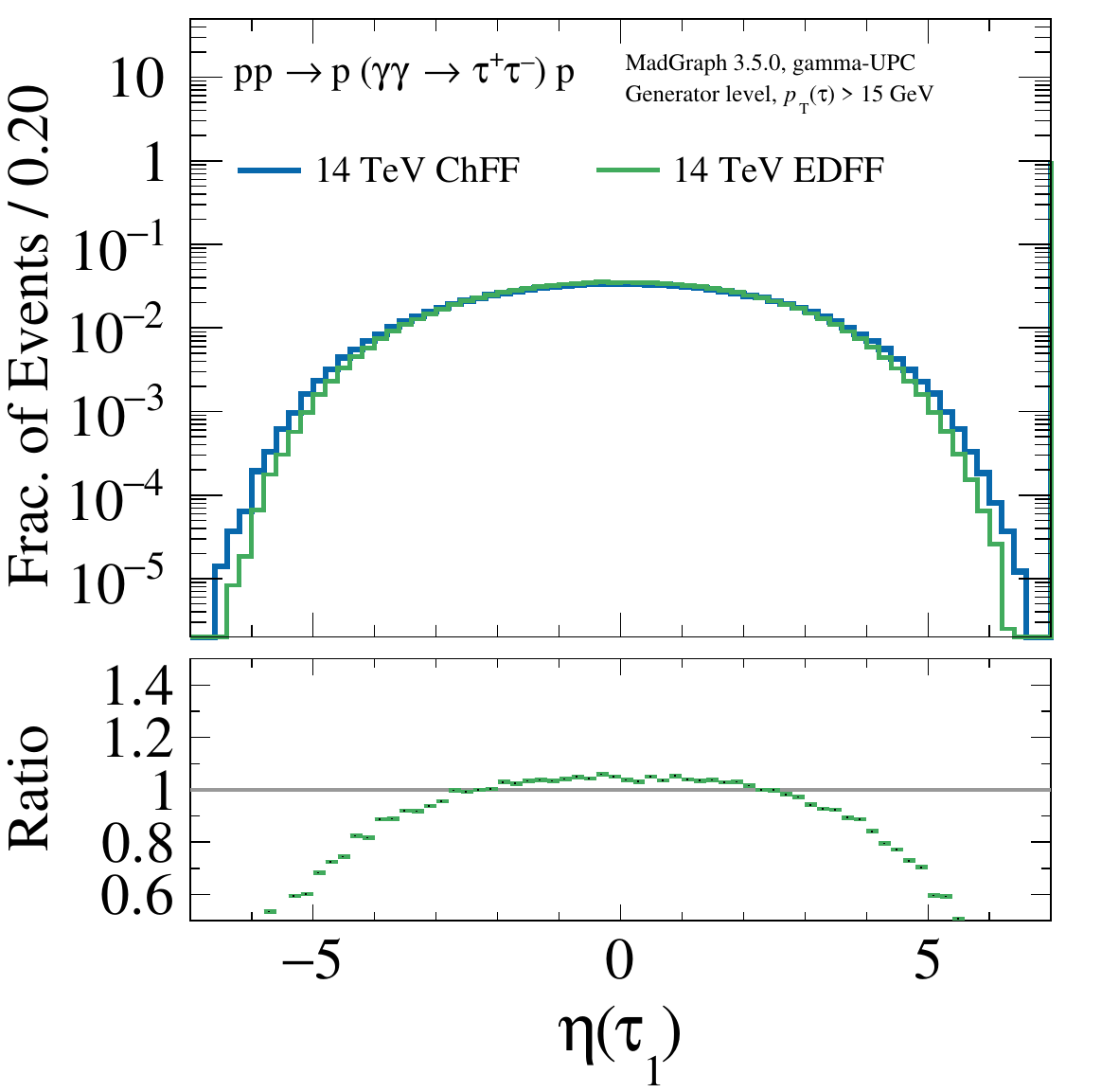}
    \caption{Unit normalized generator-level \textsc{MadGraph} distributions of $\gamma\gamma\to\tau\tau$ comparing $pp$ at 14 TeV photon flux choices of \textsc{gamma-UPC} in \textsc{MadGraph}: charged form factor ``ChFF'' (thick blue), electric dipole form factor ``EDFF'' (medium green).
    A generator-level requirement of $p_\text{T}(\tau) > 15$ GeV is imposed. 
    The lower panel shows the ratio relative to the nominal ChFF.}
    \label{fig:UnitNorm_LHE_yyflux}
\end{figure*}

To estimate theoretical uncertainties of the photon flux, Fig.~\ref{fig:UnitNorm_LHE_yyflux} shows the impact on tau-lepton kinematics when considering an alternative photon flux denoted the Electric Dipole form factor (\texttt{EDFF}) in the \textsc{gamma-UPC} package. 
In Ref.~\cite{Shao:2022cly}, the differences from the \texttt{ChFF} nominal we adopt in this paper are discussed and compared with dielectron and dimuon meausrements. 
In brief, they found the \texttt{ChFF} flux provides better differential modelling but slightly overestimates the data by around 10\% compared with the \texttt{EDFF} flux. 
We find differences of around 10 to 20\% at high $m_{\tau\tau}$ and 20 to 30\% for $p_\text{T}^\tau$ tails. 
Fortunately, experimental analyses can constraint these differences using data-driven control sample techniques as discussed in the main text. 
We expect future measurements of dilepton standard candles and theoretical work will improve these theoretical uncertainties.

\begin{figure*}
    \centering
    \includegraphics[width=0.33\textwidth]{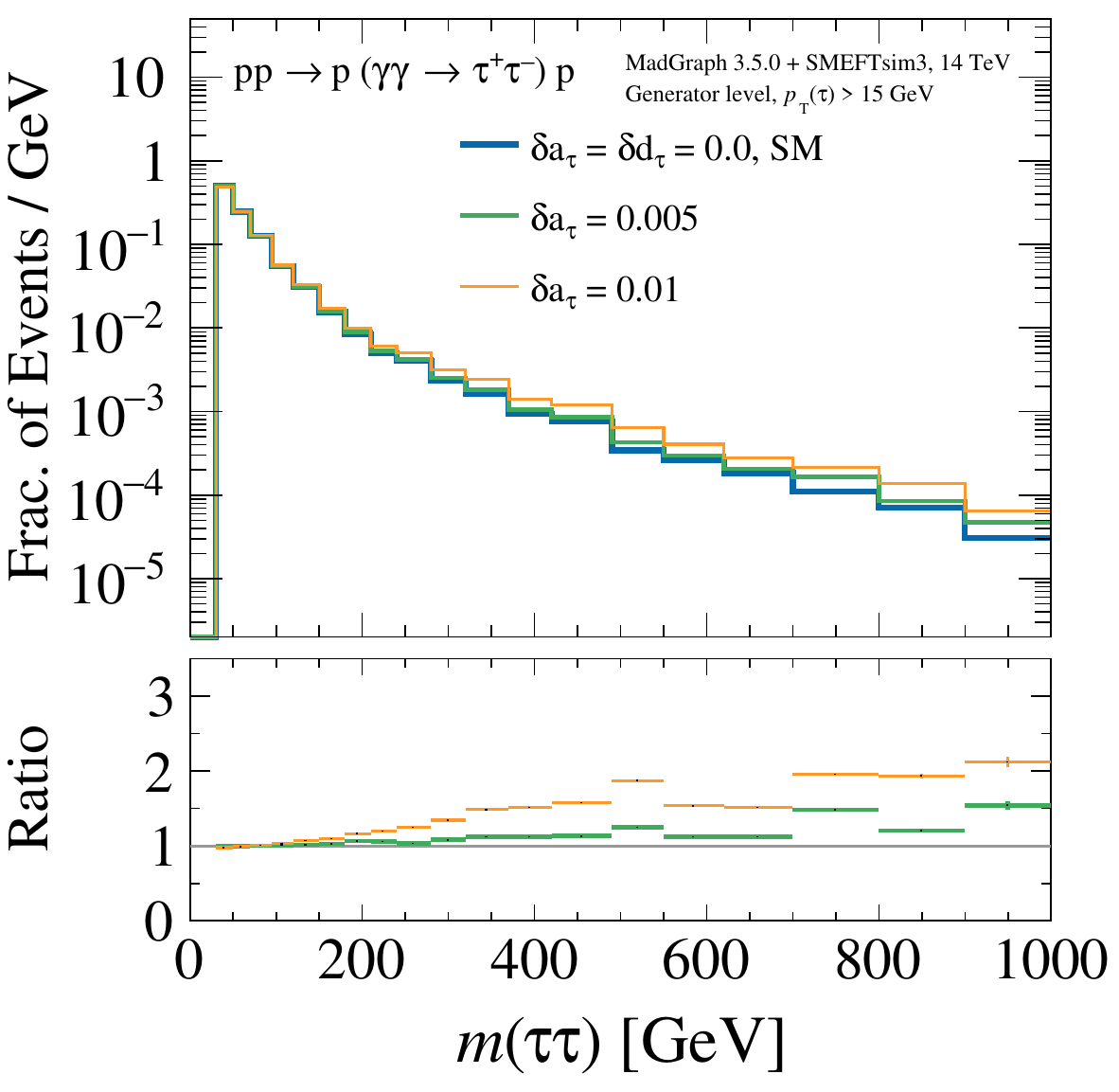}%
    \includegraphics[width=0.33\textwidth]{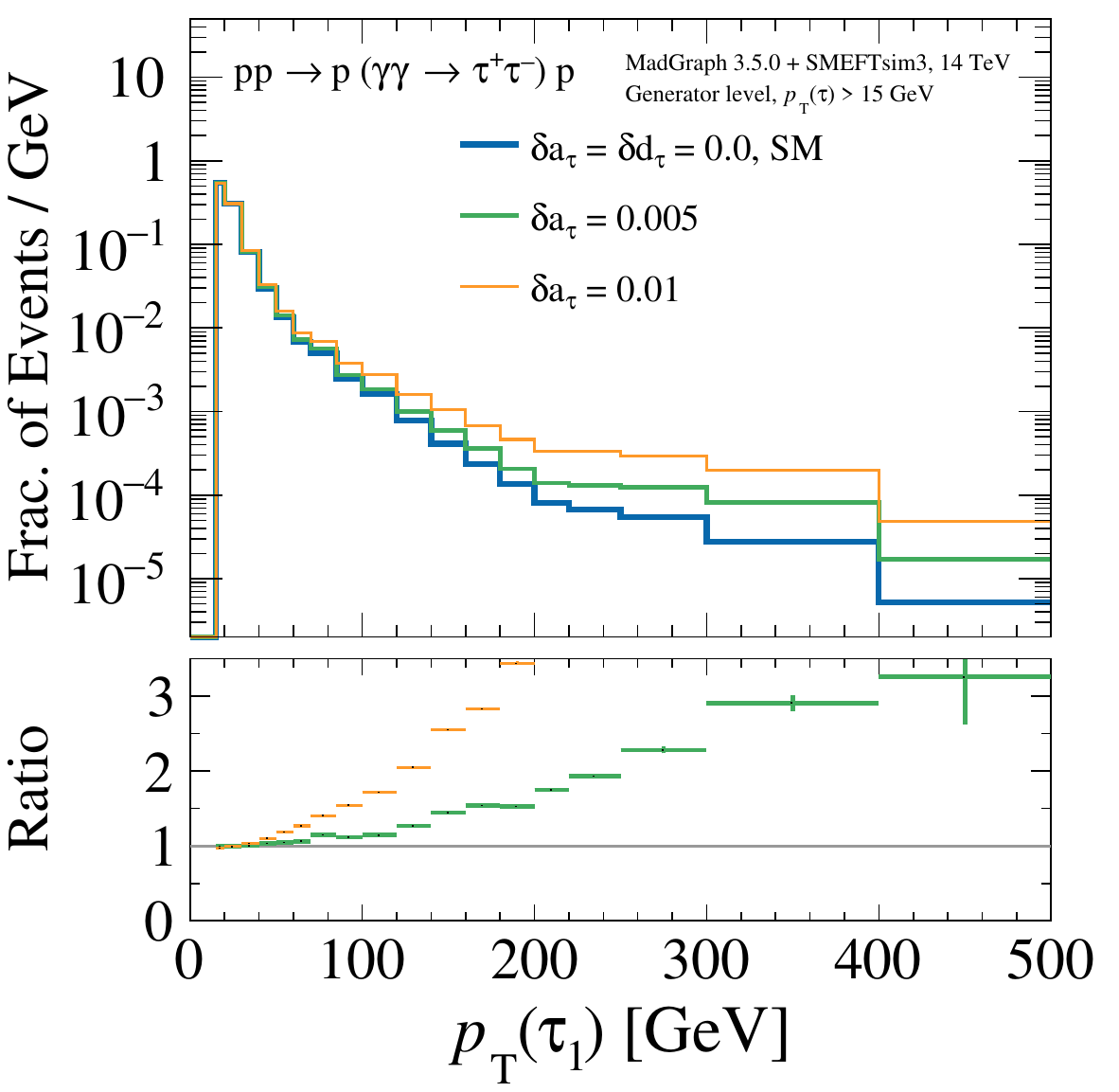}%
    \includegraphics[width=0.33\textwidth]{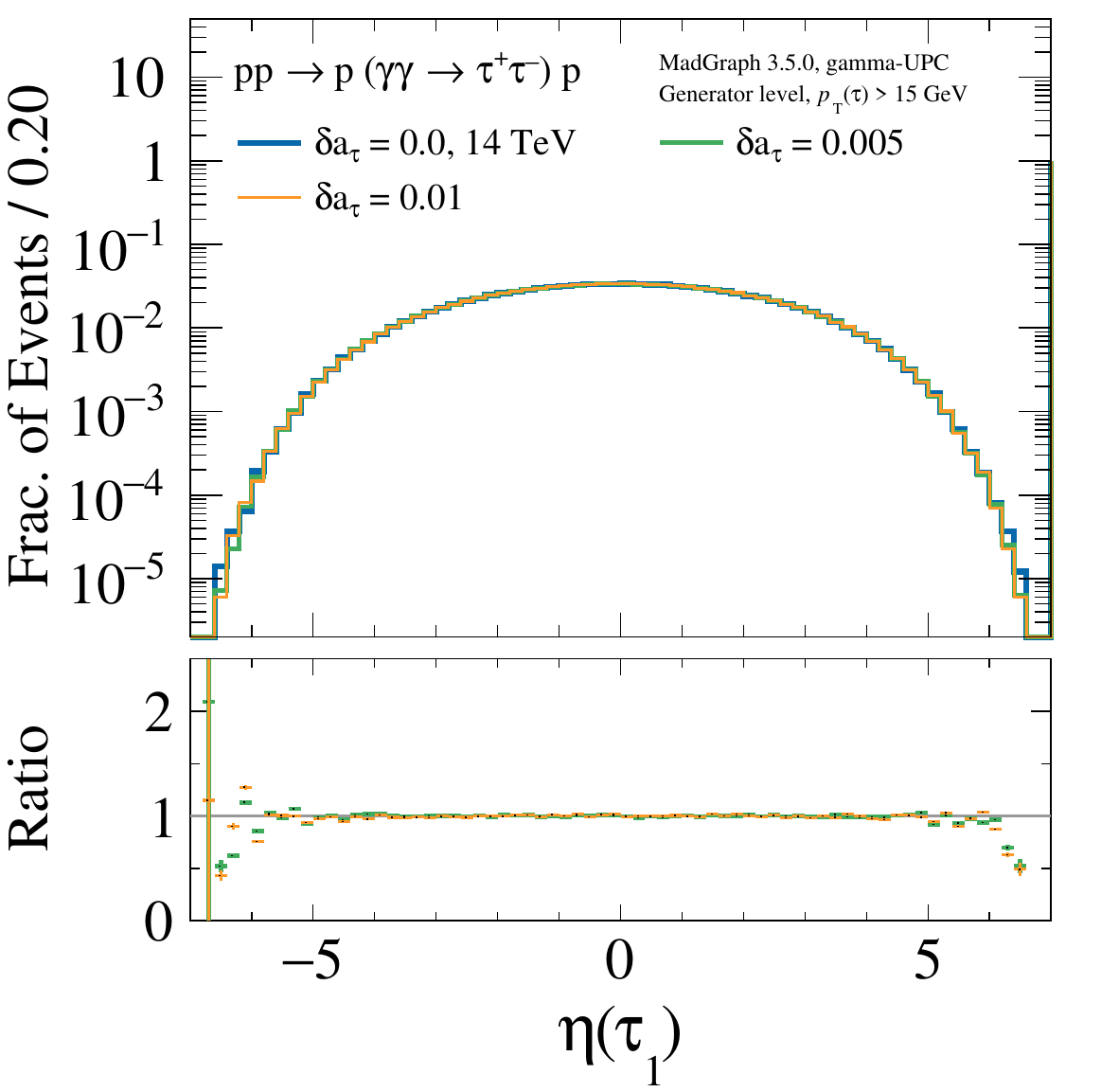}
    \caption{Unit normalized generator-level \textsc{MadGraph} distributions of $\gamma\gamma\to\tau\tau$ comparing magnetic dipole variations $\delta a_\tau$ (upper) for $pp$ at 14 TeV photon flux choices of \textsc{gamma-UPC} in \textsc{MadGraph}. 
    A generator-level requirement of $p_\text{T}(\tau) > 15$ GeV is imposed. 
    The lower panel of each subfigure shows the ratio relative to the SM nominal.}
    \label{fig:UnitNorm_LHE_varatau}
\end{figure*}

Figure~\ref{fig:UnitNorm_LHE_varatau} shows how shifts in magnetic dipole moments $\delta a_\tau$ impact tau-lepton kinematics. 
These shape differences are modest in the $m_{\tau\tau}$ and especially striking in the $p_\text{T}^\tau$ spectra, rising at high values as expected from effective field theory. 
Changes in these spectra also modify the kinematics of the tau-lepton decay products. 
Analyses can therefore fit observables such as the electron and muon $p_\text{T}^{\ell}$ spectra in $\tau \to \ell\nu\nu$ decays to substantially enhance sensitivity to $\delta a_\tau$ beyond measuring inclusive cross-sections alone, as introduced in Ref.~\cite{Beresford:2019gww}. 

\begin{figure*}
    \centering
    \includegraphics[width=0.45\textwidth]{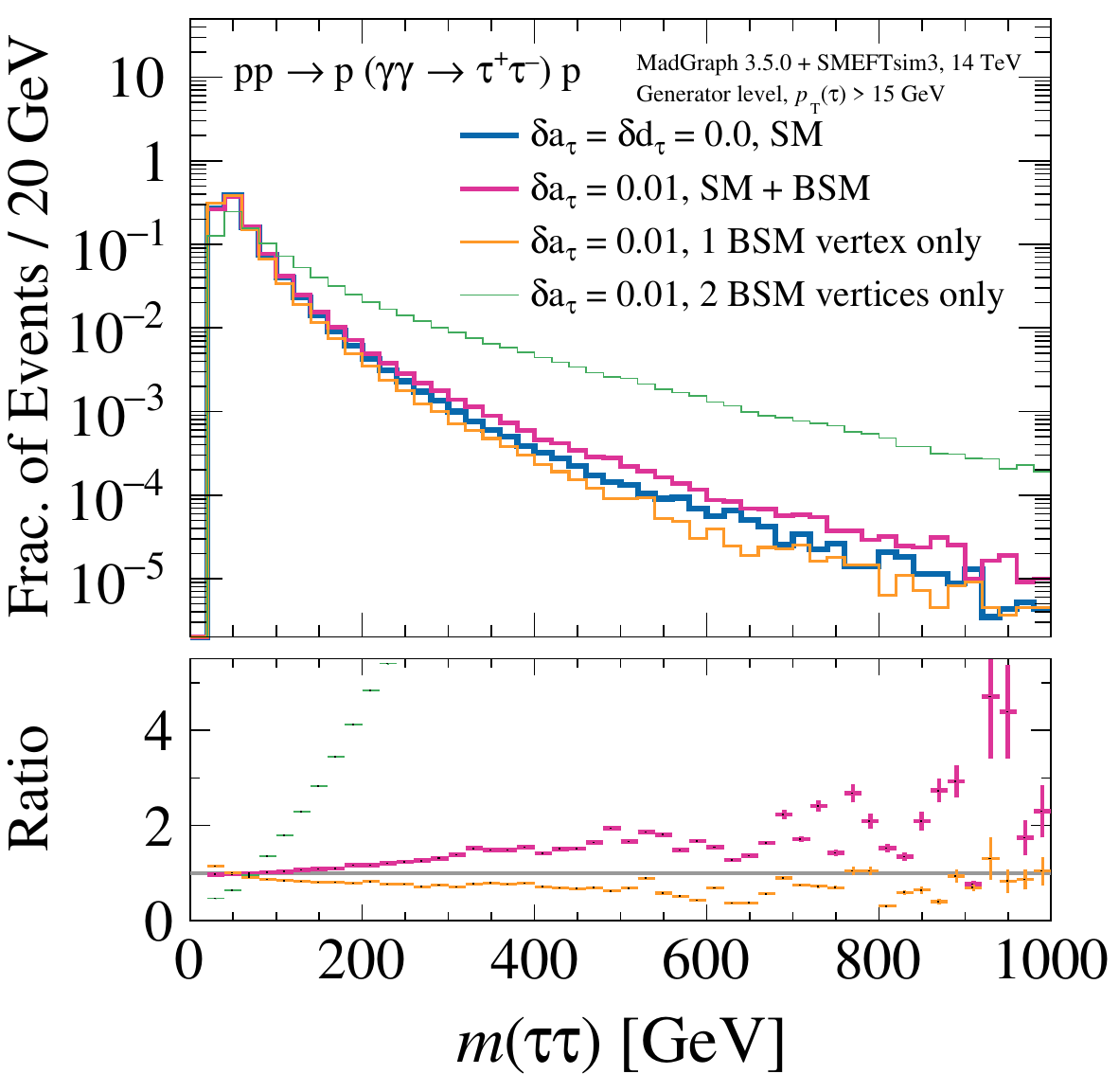}%
    \includegraphics[width=0.45\textwidth]{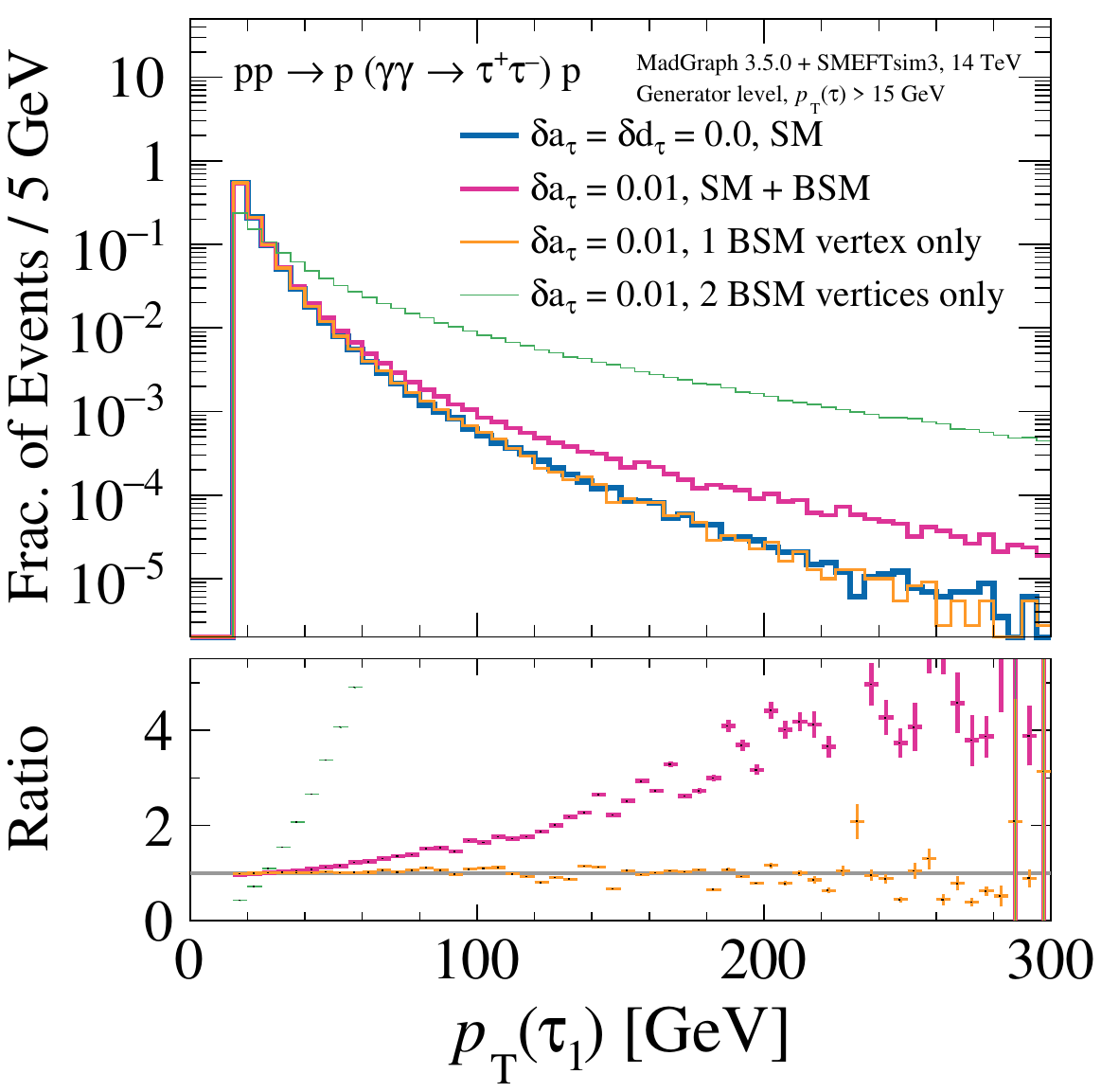}
    \caption{Unit normalized generator-level \textsc{MadGraph} distributions of $\text{pp} \to \text{p}(\gamma\gamma\to\tau\tau)\text{p}$ at 14~TeV comparing the SM-only (very thick blue), SM plus BSM (\texttt{NP\^{}2<=2}, thick pink), only 1 BSM vertex (\texttt{NP\^{}2==1}, medium orange), only 2 BSM vertices (\texttt{NP\^{}2==2}, thin green) for $\delta a_\tau = 0.01$.
    A generator-level requirement of $p_\text{T}^\tau > 15$ GeV is imposed. 
    The lower panel shows the ratio relative to the SM nominal.}
    \label{fig:UnitNorm_LHE_atauNP}
\end{figure*}

Figure~\ref{fig:UnitNorm_LHE_atauNP} compares how SM, BSM-only and their combined (SM + BSM) diagrams in the matrix elements impact unit-normalized differential distributions of $m(\tau\tau)$ and $p_\text{T}^\tau$.
To generate only the linear (quadratic) pieces with one (two) BSM dipole vertices entering the matrix element, we set \texttt{NP\^{}2==1} (\texttt{NP\^{}2==2}) in \textsc{MadGraph} .
Displayed are shifts in the magnetic dipole of $\delta a_\tau = 0.01$.

\begin{figure*}
    \centering
    \includegraphics[width=0.49\textwidth]{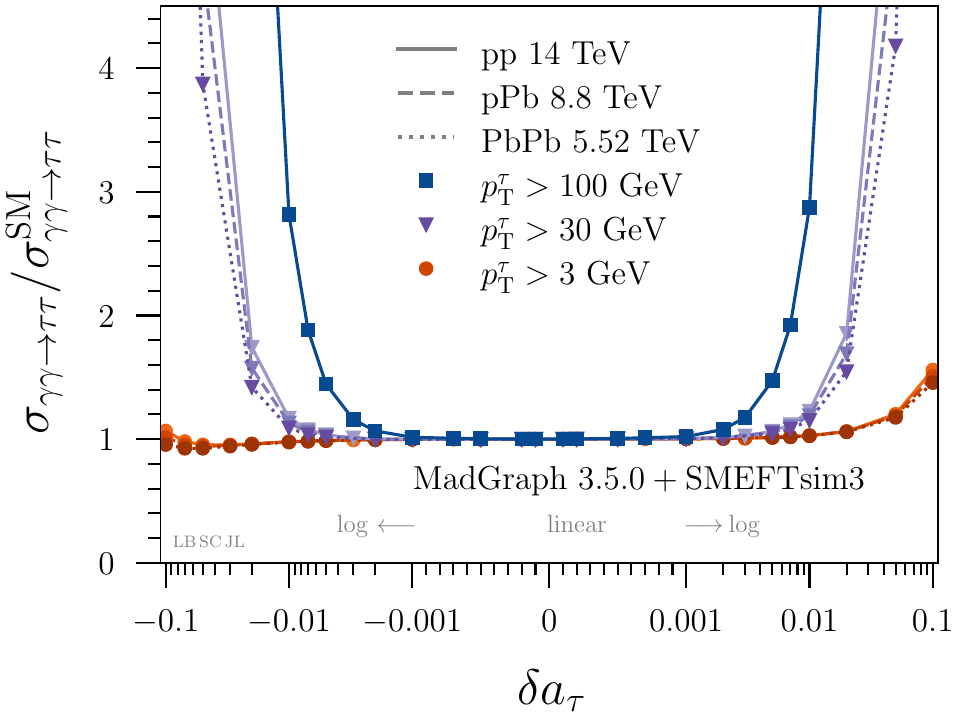}%
    \includegraphics[width=0.49\textwidth]{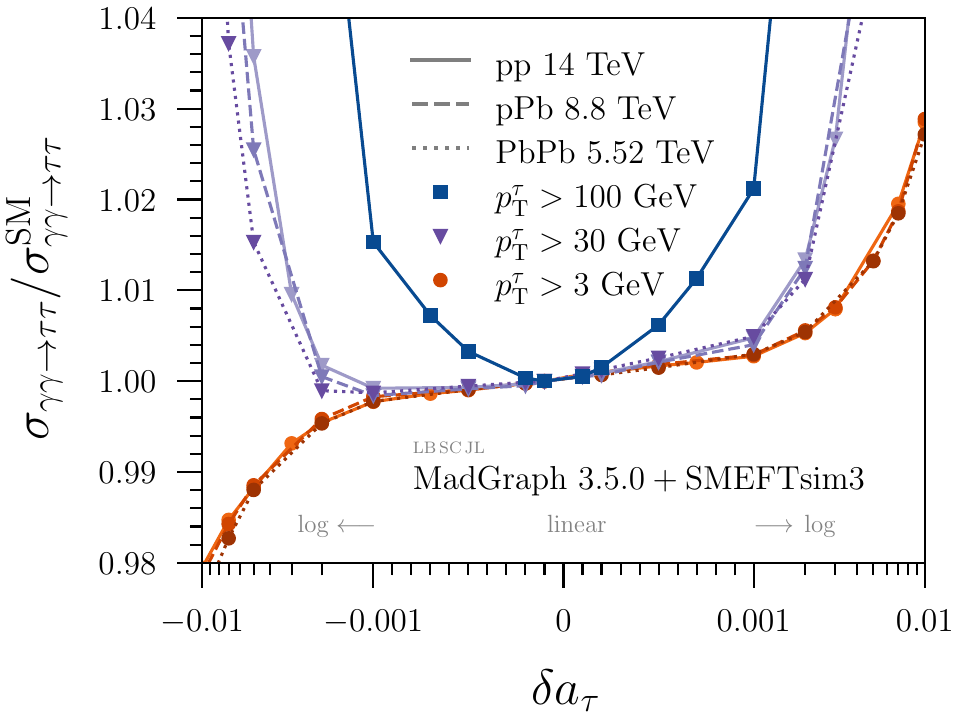}
    \caption{Wider (left) and zoomed (right) axis version of Fig.~\ref{fig:aa2tautau_xsec} showing cross-sections relative to the SM computed using \textsc{MadGraph}+\textsc{SMEFTsim} for elastic photon-fusion production of tau-lepton pairs $\sigma_{\gamma\gamma\to\tau\tau}/\sigma_{\gamma\gamma\to\tau\tau}^\text{SM}$ vs magnetic moment variations $\delta a_\tau$. 
    Various beam types and center-of-mass energies are displayed: proton-proton pp 14~TeV (solid), proton-lead pPb 8.8~TeV (dashed), lead-lead PbPb 5.52~TeV (dotted). 
    Different generator requirements on the minimum tau-lepton transverse momentum $p_\text{T}^\tau$ are shown: 100~GeV (blue squares, only pp), 30~GeV (purple triangles), 3~GeV (orange circles).
    Note the linear-log scale division at $\delta a_\tau = \pm 0.001$.}
    \label{fig:aa2tautau_xsec_large_zoom}
\end{figure*}

Figure~\ref{fig:aa2tautau_xsec_large_zoom} shows the enlarged and zoomed axis versions of Fig.~\ref{fig:aa2tautau_xsec} for the $\gamma\gamma\to\tau\tau$ cross-section varying with magnetic dipole shifts $\delta a_\tau$ assuming $\delta d_\tau = 0$. 
This highlights the dramatic impact on the cross-section variations $\sigma_{\gamma\gamma\to\tau\tau}/\sigma^\text{SM}_{\gamma\gamma\to\tau\tau}$ as we consider tighter requirements on $p_\text{T}^\tau > 3, 30, 100$~GeV, reserving the 100~GeV requirement only for pp.
By contrast, for a given minimum $p_\text{T}^\tau$, the different beam types with corresponding center-of-mass energies (pp, pPb, PbPb) have a subdominant effect.
For $p_\text{T}^\tau > 3, 30$~GeV, the cross-section changes by less than 1\% for $-0.003 < \delta a_\tau < 0.002$, whereas imposing $p_\text{T}^\tau > 100$~GeV provides a percent-level cross-section change even for shifts below per-mille $-0.0008 < \delta a_\tau < 0.0007$. 
Table~\ref{tab:aa2tautau_xsec_pb} display the absolute total LHC cross-sections of elastic $\gamma\gamma\to\tau\tau$ production when varying the magnetic $\delta a_\tau$.
This considers various beam configurations and generator requirements on the tau-lepton transverse momentum. 

\begin{table*}
\begin{center}
\renewcommand{\arraystretch}{1.2}
  \begin{ruledtabular}\begin{tabular}{cccccccc}
$\sigma^\mathrm{LHC}_{\gamma\gamma\to\tau\tau}$ [pb]: & pp 14~TeV & pPb 8.8~TeV & PbPb 5.52~TeV & pp 14~TeV & pPb 8.8~TeV & PbPb 5.52~TeV & pp 14~TeV \\ 
 $\delta a_\tau$ & $p_\mathrm{T}^{\tau} >3~\mathrm{GeV}$ & $p_\mathrm{T}^{\tau} >3~\mathrm{GeV}$ & $p_\mathrm{T}^{\tau} >3~\mathrm{GeV}$ & $p_\mathrm{T}^{\tau} >30~\mathrm{GeV}$ & $p_\mathrm{T}^{\tau} >30~\mathrm{GeV}$ & $p_\mathrm{T}^{\tau} >30~\mathrm{GeV}$ & $p_\mathrm{T}^{\tau} >100~\mathrm{GeV}$\\ 
\midrule 
0.1 & 61.27 & \num{1.499e+05} & \num{2.35e+08} & 2.181 & 2047 & \num{3.454e+05} & 0.4644 \\
0.05 & 47.23 & \num{1.181e+05} & \num{1.896e+08} & 0.628 & 610.5 & \num{1.076e+05} & 0.1179 \\ 
0.02 & 41.74 & \num{1.054e+05} & \num{1.707e+08} & 0.1901 & 202.4 & \num{3.978e+04} & 0.02109 \\ 
0.01 & 40.41 & \num{1.022e+05} & \num{1.657e+08} & 0.1256 & 142.6 & \num{2.963e+04} & 0.007192 \\ 
0.005 & 39.76 & \num{1.007e+05} & \num{1.634e+08} & 0.109 & 126.3 & \num{2.69e+04} & 0.003687 \\ 
0.002 & 39.5 & \num{9.991e+04} & \num{1.622e+08} & 0.1038 & 121.2 & \num{2.602e+04} & 0.0027 \\ 
0.001 & 39.4 & \num{9.964e+04} & \num{1.618e+08} & 0.1029 & 120.2 & \num{2.586e+04} & 0.002556 \\ 
0.0 & 39.22 & \num{9.933e+04} & \num{1.611e+08} & 0.1024 & 120 & \num{2.581e+04} & 0.002505 \\ 
$-$0.001 & 39.21 & \num{9.918e+04} & \num{1.609e+08} & 0.1023 & 119.5 & \num{2.57e+04} & 0.002541 \\ 
$-$0.002 & 39.12 & \num{9.894e+04} & \num{1.605e+08} & 0.1026 & 119.8 & \num{2.57e+04} & 0.002671 \\ 
$-$0.005 & 38.83 & \num{9.821e+04} & \num{1.594e+08} & 0.1061 & 122.8 & \num{2.613e+04} & 0.003615 \\ 
$-$0.01 & 38.49 & \num{9.723e+04} & \num{1.574e+08} & 0.1198 & 135.6 & \num{2.806e+04} & 0.007048 \\ 
$-$0.02 & 37.82 & \num{9.547e+04} & \num{1.544e+08} & 0.1784 & 188.4 & \num{3.655e+04} & 0.02079 \\ 
$-$0.05 & 37.52 & \num{9.356e+04} & \num{1.493e+08} & 0.5982 & 575.6 & \num{9.965e+04} & 0.1172 \\ 
$-$0.1 & 41.91 & \num{1.007e+05} & \num{1.542e+08} & 2.119 & 1977 & \num{3.294e+05} & 0.4627 \\ 

\end{tabular}
\end{ruledtabular}
\end{center}
\caption{ 
Generator-level LHC cross-sections $\sigma_{\gamma\gamma\to\tau\tau}^\text{LHC}$ in picobarns (pb)
for elastic photon-fusion production of tau-lepton pairs using \textsc{MadGraph}~3.5.0 and the charged form factor photon fluxes from \textsc{gamma-UPC}.
This is interfaced with the \textsc{SMEFTsim}3 package to evaluate for variations in the anomalous magnetic moment $\delta a_\tau$.
The columns show different beam species with their corresponding center-of-mass energies,
and different generator-level requirements on the tau-lepton transverse moment $p_\text{T}^{\tau}$.
}
\label{tab:aa2tautau_xsec_pb}
\end{table*} 
\begin{table}[!ht]
\begin{center}
\renewcommand{\arraystretch}{1.1}
  \begin{tabular*}{0.6\textwidth}{@{\extracolsep{\fill}}lrrrr}
  \toprule
   Requirement 
 & $qq\rightarrow\tau\tau$
 & $qq\rightarrow WW$
 & $\gamma\gamma\rightarrow WW$
 & $\gamma\gamma\rightarrow\tau\tau$
 \\
 \midrule
 $\sigma \times \textrm{BR} \times \mathcal{L} \times \varepsilon_{\textrm{veto}}$ & $92652.0$  & $7562.8$  & $23347.0$ & $226900.8$ \\
 $2\ell$ & $17008.9$  & $1057.1$  & $371.0$ & $4882.0$ \\
 $p_\textrm{T}^{e} > 18~\textrm{GeV}$, $|\eta^{e}| < 2.5$,\\$p_\textrm{T}^{\mu} > 15~\textrm{GeV}$, $|\eta^{\mu}| < 2.5$ & $2984.5$  & $628.9$  & $214.5$ & $231.6$ \\
 $m_{e\mu} > 20~\rm{GeV}$ & $2953.2$  & $610.8$  & $210.7$  & $230.6$ \\
 $|\Delta\Phi_{e\mu}| > 3.10$ & $425.4$  & $17.0$  & $6.5$ & $122.1$ \\
 $m_{T2}^{100} < 101$ & $424.8$  & $13.7$  & $4.6$ & $121.1$ \\
 \bottomrule
 \end{tabular*}
\end{center}
\caption{Cutflow of yields after each requirement applied sequentially, normalised to $\mathcal{L} = 300.0$~fb$^{-1}$. The pileup track-veto efficiency is $\varepsilon_\text{PU}=50\%$ and the underlying event efficiency is $\varepsilon_\text{UE}= 0.4\%$. A systematic uncertainty on the signal and background yields of 5\% is assumed.}
\label{tab:cutflow_main_emu}
\end{table}

\end{document}